\documentclass[sigconf]{acmart}

\usepackage[utf8]{inputenc}
\usepackage{amsmath, amsfonts}
\usepackage{paralist}
\usepackage{tikz}
\usepackage{graphicx}
\usepackage{xspace}
\usepackage{url}
\usepackage{hyperref}
\usepackage{multirow}
\usepackage{rotating}
\usepackage{makecell}
\usepackage{algorithm}
\usepackage[noend]{algpseudocode}
\usepackage{adjustbox}
\usepackage{rotating}
\usepackage{tikz}
\usetikzlibrary{positioning}
\usepackage{pdflscape}
\usepackage{hyperref}
\usepackage{subcaption}
\usepackage{endnotes}
\usepackage{longtable,booktabs}

\newcommand{\bron}{\textbf{BRON}\xspace}
\newcommand{\TACTIC}{\textit{Tactic}\xspace}  
\newcommand{\TACTICS}{\textit{Tactics}\xspace}  
\newcommand{\TECHNIQ}{\textit{Technique}\xspace}  
\newcommand{\TECHNIQS}{\textit{Techniques}\xspace}  
\newcommand{\WEAKNESS}{\textit{Weakness}\xspace}  
\newcommand{\VULNERABILITY}{\textit{Vulnerability}\xspace} 
\newcommand{\VULNERABILITIES}{\textit{Vulnerabilities}\xspace} 
\newcommand{\platform}{\textit{Known Affected Hardware or Software Configuration}\xspace}
\newcommand{\platforms}{\platform{s}\xspace}

\newcommand{\PLATFORMS}{\platforms}
\newcommand{\shortconfig}{\textit{Affected Prod Conf}}
\newcommand{\config}{\textit{Affected Product Configuration}\xspace}  
\newcommand{\APC}{\textit{APC}\xspace}
\newcommand{\configs}{\config{s}\xspace}
\newcommand{\CONFIG}{\config}  
\newcommand{\CONFIGS}{\config{\textit{s}}\xspace}
\newcommand{\AttackPattern}{\textit{Attack Pattern}\xspace}
\newcommand{\AttackPatterns}{\textit{Attack Patterns}\xspace}
\newcommand{\ATTACKPATTERN}{\textit{Attack Pattern}\xspace}
\newcommand{\ATTACKPATTERNS}{\textit{Attack Patterns}\xspace}
\newcommand{\graph}{graph\xspace} 

\newcommand{\unlinked}{non-demonstrated\xspace}  

\newcommand{\linked}{operational\xspace}
\newcommand{\Linked}{Operational\xspace}

\newcommand{\BRONdb}{\bron}
\newcommand{\brongdb}{\bron-\texttt{Graph-DB}\xspace}
\newcommand{\bronjson}{\bron-\texttt{JSON}\xspace}

\newcommand{\Bron}{\BRONdb}

\newcommand{\floater}{Floating Entry\xspace} 
\newcommand{\floaters}{Floating Entries\xspace} 

\newcommand{\supernodes}{\textit{Super Entries}\xspace}

\newcommand{\ResearchQ}{\textit{RQ:}\xspace}

\AtBeginDocument{%
  \providecommand\BibTeX{{%
    \normalfont B\kern-0.5em{\scshape i\kern-0.25em b}\kern-0.8em\TeX}}}

\setcopyright{none}

\begin{document}
%
\title{Linking Threat Tactics, Techniques, and Patterns with Defensive Weaknesses, Vulnerabilities and Affected Platform Configurations for Cyber Hunting}

\author{
{\rm Erik Hemberg}\\
MIT,
hembergerik@csail.mit.edu
\and
{\rm Jonathan Kelly} \\
MIT,
jgkelly@alumn.mit.edu
\and
{\rm Michal Shlapentokh-Rothman}\\
MIT,
mshlapen@alumn.mit.edu
\and
{\rm Bryn Reinstadler} \\
MIT,
brynr@mit.edu
\and
{\rm Katherine Xu} \\
MIT,
katexu@mit.edu
\and
{\rm Nick Rutar} \\
Perspecta Labs,
nrutar@perspectalabs.com
\and
{\rm Una-May O'Reilly} \\
MIT,
  unamay@csail.mit.edu
}


\renewcommand{\shortauthors}{A. Non, et al.}

\begin{abstract}
Many public sources of cyber threat and vulnerability information exist to help defend cyber systems. This paper links MITRE’s ATT\&CK   MATRIX of Tactics and Techniques, NIST’s  Common Weakness Enumerations (CWE),  Common Vulnerabilities and Exposures (CVE), and Common Attack Pattern Enumeration and Classification list (CAPEC), to gain further insight from alerts, threats and vulnerabilities. We preserve all  entries and relations of the sources, while enabling bi-directional, relational path tracing within an aggregate data graph called \bron. In one example, we use \bron to enhance the information derived from a list of the top 10 most frequently exploited CVEs. We identify attack patterns, tactics, and techniques that exploit these CVEs and also uncover a disparity in how much linked information exists for each of these CVEs. This prompts us to further inventory \bron's collection of sources to provide a view of the extent and range of the coverage and blind spots of public data sources. 

\end{abstract}



\maketitle

\section{Introduction}


%
%
%
%
%
%

Cyber threats are harmful and perpetrated by both state and criminal
actors. Large scale community efforts to share information about them
facilitate better security. Similarly, because networks overlap in
their deployed software and hardware, information on vulnerabilities,
affected products, exploits, and patches/fixes is often shared via
community efforts.  In this contribution, we data mine a set of these
information sources in order to expand upon their defensive utility
for threat hunting, i.e. actively exploring and pursuing potential
advanced persistent threats~\cite{milajerdi2019poirot}. These sources are all maintained by MITRE and NIST. The sources
are:
\begin{inparaenum}[\itshape 1)]
\item \texttt{MITRE ATT\&CK} \TACTICS~\cite{attack}, 
\item \texttt{MITRE ATT\&CK} \TECHNIQS~\cite{attack}, 
\item \texttt{Common Attack Pattern Enumeration and Classification}  (CAPEC)~\cite{capec},  
\item \texttt{Common Weakness Enumerations}~(CWE)~\cite{cwe},
\item \texttt{Common Vulnerability Enumerations}~(CVE)~\cite{cve}.
\end{inparaenum}

On one axis, these sources offer both abstract threat and concrete vulnerability information. For example, a CVE has concrete information about affected products while a ATTACK \TACTIC is an abstract goal of an APT, such as \texttt{Defense Evasion}.  These two types of information have complementary value when assessing a security risk and mitigating it.  On another axis, the data spans information from threat tactics to vulnerabilities, with tactics targeting vulnerabilities.  This axis offers insights starting from threat going to vulnerability or vice versa.  

We conduct two related inquiries. The first is centered on security insight exploration for threat hunting. We query the data sources to find out how it helps with enlarging the scope of knowledge around cyber threat alerts, how it informs product security evaluations, and how it identifies connections along the threat to vulnerability axis. The second follows from a finding of the first: not all the data we hope to explore for enhancement is present. The second inquiry, therefore, introspectively examines the sources as an amalgamation in order to gauge their collective extent (and consequent value), e.g. not all data sources are connected. This inventory serves to highlight ``known and not known'', though there is inherent ambiguity between what is unknown or missing.


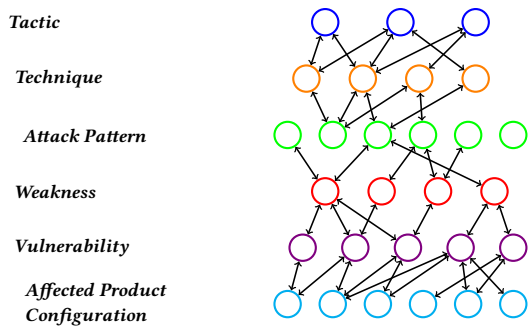
\begin{figure}[!b]
  \centering
  \scalebox{.5}{\begin{tikzpicture}[
tactic/.style={circle, draw=blue, ultra thick, minimum size = 7mm},
techniq/.style={circle, draw=orange, ultra thick, minimum size = 7mm},
capec/.style={circle, draw=green, ultra thick, minimum size = 7mm},
cwe/.style={circle, draw=red, ultra thick, minimum size = 7mm},
cve/.style={circle, draw=violet, ultra thick, minimum size = 7mm},
cpe/.style={circle, draw=cyan, ultra thick, minimum size = 7mm},
]
\node[minimum size = 10 mm] (tactic-type) at (-3.75,0) {\huge \textbf{\TACTIC}};
 \node[tactic]  (tactic1) at (4,0) {};
 \node[tactic] (tactic2) at (6,0) {};
\node[tactic] (tactic3) at (8,0) {};
\node[minimum size = 10 mm] (techniq-type) at (-3.1,-1.5) {\huge \textbf{\TECHNIQ}};
\node[techniq] (techniq1) at (3.5,-1.5) {};
\node[techniq] (techniq2) at (5,-1.5) {};
\node[techniq] (techniq3) at (6.5,-1.5) {};
\node[techniq] (techniq4) at (8.0,-1.5) {};
\draw[<->,very thick] (tactic1) -- (techniq1);
\draw[<->,very thick] (tactic1) -- (techniq2);
\draw[<->,very thick](tactic2) -- (techniq1);
\draw[<->,very thick] (tactic2) -- (techniq2);
\draw[<->,very thick] (tactic2) -- (techniq4);
\draw[<->,very thick] (tactic3) -- (techniq3);
\draw[<->,very thick] (tactic3) -- (techniq2);

\node[minimum size = 10 mm] (capec-type) at (-2.4,-3) {\huge \textbf{\ATTACKPATTERN}};
\node[capec] (capec1) at (3,-3) {};
\node[capec] (capec2) at (4.2,-3) {};
\node[capec] (capec3) at (5.4,-3) {};
\node[capec] (capec4) at (6.6,-3) {};
\node[capec] (capec5) at (7.8,-3) {};
\node[capec] (capec6) at (9.0,-3) {};

\draw[<->,very thick] (techniq2) -- (capec2);
\draw[<->,very thick] (techniq2) -- (capec3);
\draw[<->,very thick] (techniq3) -- (capec2);
\draw[<->,very thick] (techniq4) -- (capec3);
\draw[<->, very thick] (techniq3) -- (capec4);
\draw[<->, very thick] (techniq1) -- (capec2);

\node[minimum size = 10 mm] (cwe-type) at (-3.2,-4.5) {\huge \textbf{\WEAKNESS}};
\node[cwe] (cwe1) at (4.0,-4.5) {};
\node[cwe] (cwe2) at (5.5,-4.5) {};
\node[cwe] (cwe3) at (7,-4.5) {};
\node[cwe] (cwe4) at (8.5,-4.5) {};

\draw[<->,very thick] (capec1) -- (cwe1);
\draw[<->,very thick] (capec3) -- (cwe1);
\draw[<->,very thick] (capec4) -- (cwe3);
\draw[<->,very thick] (capec3) -- (cwe4);
\draw[<->,very thick] (capec4) -- (cwe2);
\draw[<->,very thick] (capec5) -- (cwe3);

\node[minimum size = 10 mm] (cve-type) at (-2.75,-6.0) {\huge \textbf{\VULNERABILITY}};
\node[cve] (cve1) at (3.4,-6) {};
\node[cve] (cve2) at (4.8,-6) {};
\node[cve] (cve3) at (6.2,-6) {};
\node[cve] (cve4) at (7.6,-6) {};
\node[cve] (cve5) at (9.0,-6) {};

\draw[<->,very thick] (cwe1) -- (cve1);
\draw[<->,very thick] (cwe1) -- (cve2);
\draw[<->,very thick] (cwe1) -- (cve3);
\draw[<->,very thick] (cwe2) -- (cve2);
\draw[<->, very thick] (cwe3) -- (cve3);
\draw[<->, very thick] (cwe4) -- (cve4);
\draw[<->, very thick] (cwe4) -- (cve5);

\node[minimum size = 10 mm, align=left] (cpe-type) at (-2.1,-7.5) {\huge \textbf{\textit{Affected}} \huge \textbf{\textit{Product}} \\[1.5mm] \huge \textbf{\textit{Configuration}}};
\node[cpe] (cpe1) at (3,-7.5) {};
\node[cpe] (cpe2) at (4.2,-7.5) {};
\node[cpe] (cpe3) at (5.4, -7.5) {};
\node[cpe] (cpe4) at (6.6,-7.5) {};
\node[cpe] (cpe5) at (7.8,-7.5) {};
\node[cpe] (cpe6) at (9.0,-7.5) {};

\draw[<->,very thick] (cve1) -- (cpe1);
\draw[<->, very thick] (cve2) -- (cpe1);
\draw[<->,very thick] (cve2) -- (cpe2);
\draw[<->, very thick] (cve3) -- (cpe2);
\draw[<->, very thick] (cve4) -- (cpe2);
\draw[<->, very thick]  (cve3) -- (cpe3);
\draw[<->, very thick]  (cve4) -- (cpe3);
\draw[<->, very thick] (cve5) -- (cpe4);
\draw[<->, very thick] (cve4) -- (cpe5);
\draw[<->, very thick] (cve4) -- (cpe6);
\draw[<->, very thick] (cve5) -- (cpe5);

\end{tikzpicture} }
  \caption{Schematic of \bron{'s} \graph of the combined sources. Source entries are nodes and relational links are edges. In the original source most types of links are uni-directional.  The \graph facilitates search in the opposite direction of the source links.}
  \label{fig:simple_schema}
\end{figure}

 \begin{table*}[!h]
  \small
\centering
\caption{\bron information sources and types, organization and short descriptions.} \label{tab:acro}
\begin{tabular}{p{7cm}lp{8cm}}
\textbf{Source and Type of Entry} & \textbf{Organization} & \textbf{Description} \\ \hline \hline
ATT\&CK \textbf{\TACTICS}~\cite{attack} & MITRE & 12 common tactics of attack staging. The columns of the ATT\&CK matrix. \\ \hline
ATT\&CK \textbf{\TECHNIQS}~\cite{attack} & MITRE & Means of achieving a tactical objective, organized by \TACTIC, the row elements of the ATT\&CK matrix.  In \bron, a \TECHNIQ is a child of a \TACTIC. \\ \hline
CAPEC \textbf{\ATTACKPATTERN{s}} ~\cite{capec}  & MITRE & A naming of a relational concept linking to a \TECHNIQ (parent) and/or \WEAKNESS (child). Relates to abstract how and why (\TACTIC and \TECHNIQ) of an attack objective and/or to abstract ``where''  (\WEAKNESS) target of the attack.  \\ \hline

NVD CWE Common \textbf{Weakness} Enumeration~\cite{cwe} & NIST & Security-related flaws in architecture, design, or code. \\ \hline

NVD CVE Common \textbf{Vulnerabilities and Exposures }~\cite{cve} & NIST & Security-related flaws in software and applications. \\ \hline

NVD CVE entry field: \textbf{\platform} & NIST & Specific software application or hardware platform releases that are affected, given a parent \VULNERABILITY.  Abbreviated herein as \config. We follow the Common Platform Enumeration (CPE) naming specification for \configs~\cite{cpe} that is used in this field.  \\
\hline
NVD CVE entry field: Severity score \cite{veris} & NIST & Severity is scored using the Common Vulnerability Scoring System (CVSS). The higher the score, the greater the impact of a \VULNERABILITY. CVSS scores range from 0 to 10.\\

\end{tabular}
\end{table*}

In proceeding, we combine our selected sources within a graph framework we call \Bron, see Figure~\ref{fig:simple_schema}.  \Bron provides the convenience of unification because the sources are independent but have relational links between them in 3 pairings:  ATT\&CK techniques are linked to \ATTACKPATTERN{s},  \ATTACKPATTERN{s} are linked to CWE \WEAKNESS{es}, and \WEAKNESS{es} are linked to CVE \VULNERABILITIES. \Bron has bi-directional links that simplify traversing from tactic to affected product (or vice versa) and that simplify starting from an \ATTACKPATTERN  and finding both the ATT\&CK \TECHNIQ and \WEAKNESS it links.  These links improve technical efficiency when scripting queries and running on the graph rather than the sources independently.  Each layer of the \graph denotes a different source.  

Nodes within a layer are entries from the source. For example, the second top-most layer has \TECHNIQS, and the second from the bottom has \VULNERABILITIES.
Edges in the \graph correspond to relational links between entries from different sources. While these links are not bi-directional in the sources, they can be traversed bi-directionally in \bron's \graph.

The contributions of this paper are:
\begin{asparaitem}
\item Investigations of:
  \begin{inparaenum}[\itshape a)]
    \item expanded abstract knowledge concerning alerts,
    \item queries that access all the information to provide more context,
    \item queries that link threats and vulnerabilities.
  \end{inparaenum}
\item Analysis of \Bron query performance
\item An inventory of the extent of the public sources and their inter-connections.
\item Open source software to create a graph database of \bron, including \texttt{Jupyter} notebooks for analyses.
\end{asparaitem}

Section~\ref{sec:methods} describes the data
sources. Section~\ref{sec:brondb} describes how we assembled \bron and
use it for analyses. Section~\ref{sec:experiments} is the
security-insight threat hunting 
inquiry. Section~\ref{sec:RelationalAnalysis} is the
inventory-driven inquiry. Section~\ref{sec:bron-impl-perf} is the analysis of \bron
query performance. Section~\ref{sec:related-work} situates
\bron, as a conceptual approach, within related
work. Section~\ref{sec:discussion} is a discussion.  Finally,
Section~\ref{sec:future-work} presents potential future work.

\section{Data Sources}
\label{sec:methods}


\label{sec:link-attack-tact}

Table~\ref{tab:acro} presents the information sources and two fields that are combined with \bron.  There is documentation available for all sources
online, see \cite{attack,capec,cwe,cve,cpe}. The NIST National Vulnerability Database (NVD) provides information on security-related software flaws, related product configurations, and impact metrics. Both sources -- the \texttt{Common Weakness Enumerations}~(CWE)~\cite{cwe} and \texttt{Common Vulnerability Enumerations}~(CVE)~\cite{cve} -- are built upon and fully synchronized with lists maintained by MITRE.
\begin{asparaitem}
\item The MITRE ATT\&CK matrix~\cite{attack,strom,strom2018mitre} abstractly describes cyber attack \TECHNIQS organized across twelve sequenced \TACTICS.


\item The CWEs are a community-developed source of software and hardware \textit{weakness} types.  ``\textit{Weaknesses} are flaws, faults, bugs, and other errors in software and hardware design, architecture, code, or implementation that if left unaddressed could result in systems, networks, and hardware being vulnerable to attack.''~\cite{cweFAQ}

\item The CVE list holds publicly known cybersecurity vulnerabilities. In it  a \VULNERABILITY is defined as a ``weakness in the computational logic (e.g. code) found in software and hardware components that, when exploited, results in a negative impact to confidentiality, integrity, or availability.''~\cite{nvd}. Each CVE entry, i.e. \VULNERABILITY, contains an  identification number, a description, and at least one reference for publicly known cybersecurity vulnerabilities. Additional entry information can include fix information, severity scores and impact ratings according to the Common Vulnerability Scoring System (CVSS),  and links to exploit and advisory information~\cite{cveAndNVD}.

\item A CWE entry has a relational link to a CVE entry.
The relationship implies that the \VULNERABILITY  is an example of the (type of) \WEAKNESS.
The CVE entry  provides a field with severity score in CVSS format and \platforms. These \configs (or \shortconfig{s}, our abbreviation) document specific software or hardware releases that are affected.
\configs specifically are of interest because, with an inventory, security operators and analysts can scan them to be alerted to specific targets in their systems.

\item The CAPEC list enumerates and classifies \ATTACKPATTERNS to support identifying and understanding attacks.
\ATTACKPATTERNS  connect the ATT\&CK matrix to the CWE source, functioning as bridges that link a  \TECHNIQ within a \TACTIC to a CWE entry, i.e. \WEAKNESS. They tie a means of attack (serving a specific tactical objective, i.e. \TACTIC) to its targeted weakness.
Since a \WEAKNESS in CWE can link to a \VULNERABILITY  in CVE, and that entry can link to specific \platforms, \ATTACKPATTERNS with this ``end-to-end'' linkage are means of connecting all the way from a \TACTIC that an attacker may choose to specific \platforms that could be affected.
Vice versa, given some \config, these \ATTACKPATTERNS make it possible to see what \TACTIC and \TECHNIQS an adversary could use to target it.  As the forthcoming inventory shows, \ATTACKPATTERNS often lack links to either a \WEAKNESS or a \TECHNIQ.
\end{asparaitem}

In an example of pairwise linkages, a \textit{Discovery} \TACTIC with a \TECHNIQ
named \texttt{System Network Configuration Discovery} links to
\texttt{CAPEC-309}, \texttt{Network Topology Mapping}
which is related to \WEAKNESS \texttt{CWE-200}, \texttt{Exposure of Sensitive Information to an Unauthorized Actor}. For this \WEAKNESS there are 6,624 \VULNERABILITIES such as \texttt{CVE-2018-8433}, \texttt{Microsoft Graphics Component Information Disclosure Vulnerability}. \texttt{CVE-2018-8433} is linked to 15 \CONFIGS.


\section{\Bron}\label{sec:brondb}

Figure~\ref{fig:simple_schema} schematically depicts the selected sources as they are represented in \bron.    \bron is a relational graph which represents the entries of its different information sources as specific types of nodes and their internal and external linkages as edges.
Links that are unidirectional in the sources are tagged as such and represented bi-directionally in \bron's \graph. A \bron \graph node has properties which are the (verbatim) information on the specific entry. Node properties have been standardized across the different entry types for ease of analysis and extension, e.g. we maintain the CPE notation for \config. \bron is populated by downloading data from the data sources described in Table~\ref{tab:acro}.  

To reference \bron's information is to therefore query a graph. For example, to  quantify relationships, we trace edges and count them and/or nodes.
Since it uses a bi-directional \graph, \bron offers path finding all the way from abstract attacker tactical goals, such as \texttt{Persistence}, down to the specific applications that can be targeted as part of an attack, or vice versa. This can be done with its information sources ``in the wild'' but less conveniently. \bron essentially eliminates the need for enumerative search needed to invert a directional link.  \bron offers direct searching capability from any node, and it provides traversal along any edge relation, starting and ending from any two entries of different sources.  

For the data mining in the following sections, we use the data
available from the information sources in Table~\ref{tab:acro} as of
March 2020 and updated only the CVE data as of June 2020 to populate
\bron.

\section{Exploring \bron For Security Insights}\label{sec:experiments}\label{sec:meta-anal}

%

A cyber threat hunter turns the predator into prey by actively
exploring and pursuing potential advanced persistent threats assumed to already be in
their cyber environment ~\cite{ibm2017wp,cyberReason,milajerdi2019poirot}.  To proceed, a hunter initially asks the question:  “Am I under attack?’'~\cite{cyberReason}
To respond, they must hypothesize where they would find a threat,
when it could have happened, who may lie behind it and what techniques it uses.  This poses a ``needle-in-a-haystack'-like problem
so  good cyber hunting practice treats the hunt as a scientific
experiment where data is gathered to
prove or disprove the hypothesis~\cite{redcanaryCH}. The ATT\&CK framework, only one part of \Bron, as a knowledge-base of adversary behavior,  provides a compendium of what adversaries in general are doing but it doesn't indicate what a hunter should choose to look for. The hunter, only once they have determined what's important to look for, uses ATT\&CK~\cite{redcanaryCH}.

\bron expands on the capabilities of ATT\&CK to aid this hypothesis-driven threat
hunting in two ways. First, if a hypothesis starts from a specific APT or goal at the \TACTIC-level, it extends ATT\&CK knowledge on \TECHNIQ to \AttackPatterns (CAPEC) that connect to software weaknesses and applications with specific vulnerabilities or exposures.  For example, per Section~\ref{sec:persistenceToChrome}, the hypothesis of “I could be under a persistence-establishing attack”  can be refined all the way down to ``More specifically, a persistence-establishing attack could be coming through our Chrome browsers''. Because this type of hypothesis starts with ATT\&CK information, which we show at the top of \bron graph visualizations, we call this \bron's top-down hunting support.

Second, \bron supports a hypothesis starting from a potential target. An example is ``Our Chrome browsers are under attack and this will lead to privilege escalation''.  The hunter starts with CVE entry search, looking for instance of Chrome browser vulnerabilities. Knowledge from within CVE entries that list Chrome browser vulnerabilities potentially allows the hunter to take some actions, perhaps checking what versions are on their system. Then, the knowledge within CWE entries linked to these same CVE entries, informs them of how to leverage  tools and logging to give visibility into activities that reveal privilege escalation. As a final step, links from the CWE entry to \AttackPattern (CAPEC) and \TECHNIQ informs them, e.g. to look at activity involving files that would be dangerous in the wrong hands (access control). Because this type of hypothesis starts with \VULNERABILITY information, which we show at the bottom of \bron graph visualizations, we call this \bron's bottom-up hunting support.


In Section~\ref{sec:persistenceToChrome} we show how \bron can help
during threat hunting and provide context to a
threat, both with top-down and bottom-up hunting support. Section~\ref{sec:vend-their-brows} describes a wider threat
context for web browsers. Section~\ref{sec:alerts} analyzes properties
of alerts about common \VULNERABILITIES and \WEAKNESS{es}.

%

%

\subsection{Threat Tactics and Affected Products}\label{sec:persistenceToChrome}

Here we presume a hunter acknowledges that the network perimeter has been compromised. They ask how can they identify whether a hosted product is  targeted by a specific APT tactic, such as persistence, and vulnerable because of a particular weakness? What \AttackPattern spans the tactic and vulnerabilities?

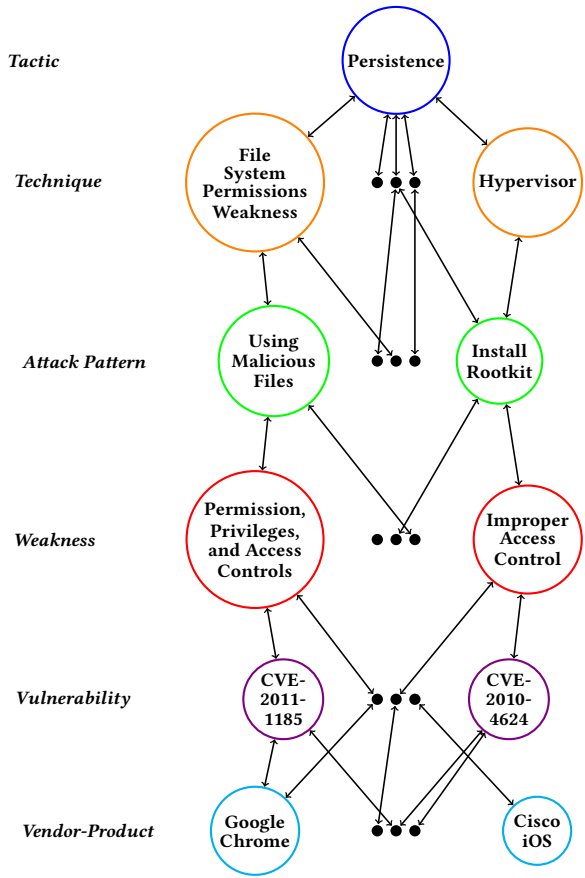
\begin{figure}[!tb]
 \centering
   \scalebox{.5}{\begin{tikzpicture}[
tactic/.style={circle, draw=blue, ultra thick, minimum size = 3mm, align=center, font=\scriptsize},
techniq/.style={circle, draw=orange, ultra thick, minimum size = 3mm, align=center, font=\scriptsize},
capec/.style={circle, draw=green, ultra thick, minimum size = 3mm, align=center, font=\scriptsize},
cwe/.style={circle, draw=red, ultra thick, minimum size = 3mm, align=center, font=\scriptsize},
cve/.style={circle, draw=violet, ultra thick, minimum size = 3mm, align=center, font=\scriptsize},
cpe/.style={circle, draw=cyan, ultra thick, minimum size = 3mm, align=center,font=\scriptsize},
dots/.style={circle, fill=black, ultra thick, minimum size = 3mm, align=center,font=\scriptsize},
]
\node[minimum size = 3 mm, align=left] (tactic-type) at (-2.65,-0.25) {\huge \textbf{\TACTIC}};
 \node[tactic]  (tactic1) at (7,-0.25) {\huge \textbf{Persistence}};

\node[minimum size = 3 mm, align=left] (technique-type) at (-2,-3.5) {\huge \textbf{\TECHNIQ}};
\node[techniq] (techniq1) at (3.25,-3.5) { \huge \textbf{File} \\[1mm] \huge \textbf{System} \\ \huge \textbf{Permissions} \\[1.5mm] \huge \textbf{Weakness}};
\node[dots] (techniqdot1) at (6.5,-3.5) {};
\node[dots] (techniqdot2) at (7,-3.5) {};
\node[dots] (techniqdot3) at (7.5,-3.5) {};
\node[techniq] (techniq3) at (10.5,-3.5) {\huge \textbf{Hypervisor}};
\draw[<->,very thick] (tactic1) -- (techniq1);
\draw[<->,very thick](tactic1) -- (techniqdot1);
\draw[<->,very thick] (tactic1) -- (techniqdot2);
\draw[<->,very thick] (tactic1) -- (techniqdot3);
\draw[<->,very thick] (tactic1) -- (techniq3);
%
%
%
\node[minimum size = 3 mm, align=left] (capec-type) at (-1.3,-8.25) {\huge \textbf{\ATTACKPATTERN}};
\node[capec] (capec1) at (3.75,-8.25) { \huge \textbf{Using} \\ \huge \textbf{Malicious} \\[1.5mm] \huge \textbf{Files}};
\node[dots] (capecdot1) at (6.5,-8.25) {};
\node[dots] (capecdot2) at (7,-8.25) {};
\node[dots] (capecdot3) at (7.5,-8.25) {};
\node[capec] (capec3) at (9.75,-8.25) {\huge \textbf{Install} \\[1.5mm] \huge \textbf{Rootkit}};
%
\draw[<->,very thick] (techniq1) -- (capec1);
\draw[<->,very thick] (techniqdot2) -- (capec3);
\draw[<->,very thick] (techniq3) -- (capec3);
\draw[<->, very thick] (techniq1) -- (capecdot2);
\draw[<->, very thick] (techniqdot2) -- (capecdot1);
\draw[<->, very thick] (techniqdot3) -- (capecdot3);
%
%
\node[minimum size = 3 mm, align=left] (cwe-type) at (-2.1,-13) {\huge \textbf{\WEAKNESS}};
\node[cwe] (cwe1) at (3.25,-13) {\huge \textbf{Permission,} \\[1mm] \huge \textbf{Privileges,} \\ \huge \textbf{and} \huge \textbf{Access} \\[1.5mm] \huge \textbf{Controls}};
\node[dots] (cwedot1) at (6.5,-13) {};
\node[dots] (cwedot2) at (7,-13) {};
\node[dots] (cwedot3) at (7.5,-13) {};
\node[cwe] (cwe3) at (10.5, -13) { \huge \textbf{Improper} \\ \huge \textbf{Access} \\[1.5mm] \huge \textbf{Control}};
%
\draw[<->,very thick] (capec1) -- (cwe1);
\draw[<->, very thick] (capec1) -- (cwedot3);
\draw[<->, very thick] (capec3) -- (cwedot2);
\draw[<->,very thick] (capec3) -- (cwe3);
%
%
%
\node[minimum size = 3 mm, align=left] (cve-type) at (-1.6,-17.25) {\huge \textbf{\VULNERABILITY}};
\node[cve] (cve1) at (4,-17.25) { \huge \textbf{CVE-} \\[1.5mm] \huge \textbf{2011-} \\[1.5mm] \huge \textbf{1185}};
\node[dots] (cvedot1) at (6.5,-17.25) {};
\node[dots] (cvedot2) at (7,-17.25) {};
\node[dots] (cvedot3) at (7.5,-17.25) {};
\node[cve] (cve3) at (10,-17.25) { \huge \textbf{CVE-} \\[1.5mm] \huge \textbf{2010-} \\[1.5mm] \huge \textbf{4624}};
%
\draw[<->,very thick] (cwe1) -- (cve1);
\draw[<->,very thick] (cwe3) -- (cve3);
\draw[<->,very thick] (cwe1) -- (cvedot1);
\draw[<->, very thick] (cwe3) -- (cvedot2);
%
%
%
\node[minimum size = 3 mm, align=left] (cpe-type) at (-1.2,-20.75) {\huge \textbf{\textit{Vendor-Product}}};
\node[cpe] (cpe1) at (3.25,-20.75) { \huge \textbf{Google} \\ \huge \textbf{Chrome}};
\node[dots] (cpedot1) at (6.5, -20.75) {};
\node[dots] (cpedot2) at (7,-20.75) {};
\node[dots] (cpedot3) at (7.5,-20.75) {};
\node[cpe] (cpe3) at (10.75,-20.75) {\huge \textbf{Cisco} \\[1.5mm] \huge \textbf{iOS}};
%
\draw[<->,very thick] (cve1) -- (cpe1);
\draw[<->,very thick] (cve1) -- (cpedot2);
\draw[<->,very thick] (cvedot2) -- (cpedot1);
\draw[<->,very thick] (cve3) -- (cpedot2);
\draw[<->,very thick] (cve3) -- (cpedot3);
\draw[<->,very thick] (cvedot1) -- (cpe1);
\draw[<->,very thick] (cvedot3) -- (cpe3);

\end{tikzpicture}}
  \caption{Subgraph of \bron showing some of the paths related to the  \TACTIC \texttt{Persistence}.
}
  \label{fig:persistence}
\end{figure}

To help address this question, we investigated the longest paths within \bron: those that run between a \TACTIC and \config.   It took advantage of \bron's source unification and bi-directional linking of these sources to find them seamlessly.  We chose a common web browser Google Chrome.  Figure~\ref{fig:persistence}  shows some of the paths that link Persistence to CVEs that list Google Chrome as an affected product.

One example of entries in  a path to Google Chrome is:
\begin{asparadesc}

\item [\TACTIC] \texttt{(TA0003)}  \texttt{Persistence}: The adversary is trying to maintain a foothold.  Persistence consists of techniques that adversaries use to keep access to systems across restarts, changed credentials, and other interruptions that could cut off access. Techniques used for persistence include any access, action, or configuration changes that let them maintain their foothold on systems, such as replacing or hijacking legitimate code or adding startup code.

\item [\TECHNIQ] \texttt{(T1574) Hijack Execution Flow}: Adversaries may execute their payloads by hijacking the way operating systems run programs. Hijacking execution flow can be for the purposes of persistence, since this hijacked execution may reoccur over time. Adversaries may also use these mechanisms to elevate privileges or evade defenses, such as application control.

\begin{asparadesc}
\item [\textit{Sub-}\TECHNIQ] \texttt{(T1574.010)} \texttt{Services File Permissions Weakness}: Adversaries may execute their own malicious payloads by hijacking the binaries used by services. Adversaries may use flaws in the permissions of Windows services to replace the binary that is executed upon service start. These service processes may automatically execute specific binaries as part of their functionality or to perform other actions. If permissions on the file system directory containing a target binary or permissions on the binary itself are improperly set, then the target binary may be overwritten with another binary using user-level permissions and executed by the original process. If the original process and thread are running under a higher permissions level, then the replaced binary will also execute under higher-level permissions, which could include SYSTEM.
\end{asparadesc}
\item  [\ATTACKPATTERN] \texttt{(CAPEC-17) Using Malicious Files}: An attack of this type exploits a system's configuration that allows an attacker to either directly access an executable file, for example through shell access; or in a possible worst case allows an attacker to upload a file and then execute it. Web servers, ftp servers, and message oriented middleware systems which have many integration points are particularly vulnerable, because both the programmers and the administrators must be in sync regarding the interfaces and the correct privileges for each interface.

\item [\WEAKNESS] \texttt{(CWE-264) Permissions, Privileges, and Access Controls}
\item [\VULNERABILITY] \texttt{CVE-2011-1185}
\begin{asparadesc}
\item [\platform] CPE is\\ { \small  \texttt{cpe:2.3:a:google:chrome:*:*:*:*:*:*:*:*}}, up to (excluding) version 10.0.648.127.
\item [\textit{Vendor-Product}] extracted as 3rd and 4th fields of \platform, obtaining Google Chrome.
\end{asparadesc}
\end{asparadesc}

A top-down interpretation of this path is:  Given an attack's objective is \texttt{Persistence}, an attack  \texttt{Using Malicious Files}, by  means of exploiting  a \texttt{Services File Permissions Weakness} \TECHNIQ, could be used to hijack execution flow and run a malicious binary due to  \texttt{Permissions, Privileges, and Access Controls} weaknesses in all versions of Google Chrome ``before 10.0.648.127''. These  ``do not prevent (1) navigation and (2) close operations on the top location of a sandboxed frame".

Bottom-up, the interpretation of this path is: If any of a given network's computers are running Google Chrome versions before 10.0.648.127, the administrators need to be on alert for the browser being hijacked to execute a malicious payload that can achieve persistence by exploiting services file permission weaknesses that allow the attack to run at a higher permission level where they can replace legitimate binaries with malicious ones.

\subsection{Web Browser threat analysis}
\label{sec:vend-their-brows}

When evaluating the security risk of different browsers, how are different ones vulnerable to threats? That is, how will they be attacked (via what \TECHNIQS) and for what actors' purposes?

\begin{figure}[!tb]
  \centering
  \includegraphics[width=0.49\textwidth]{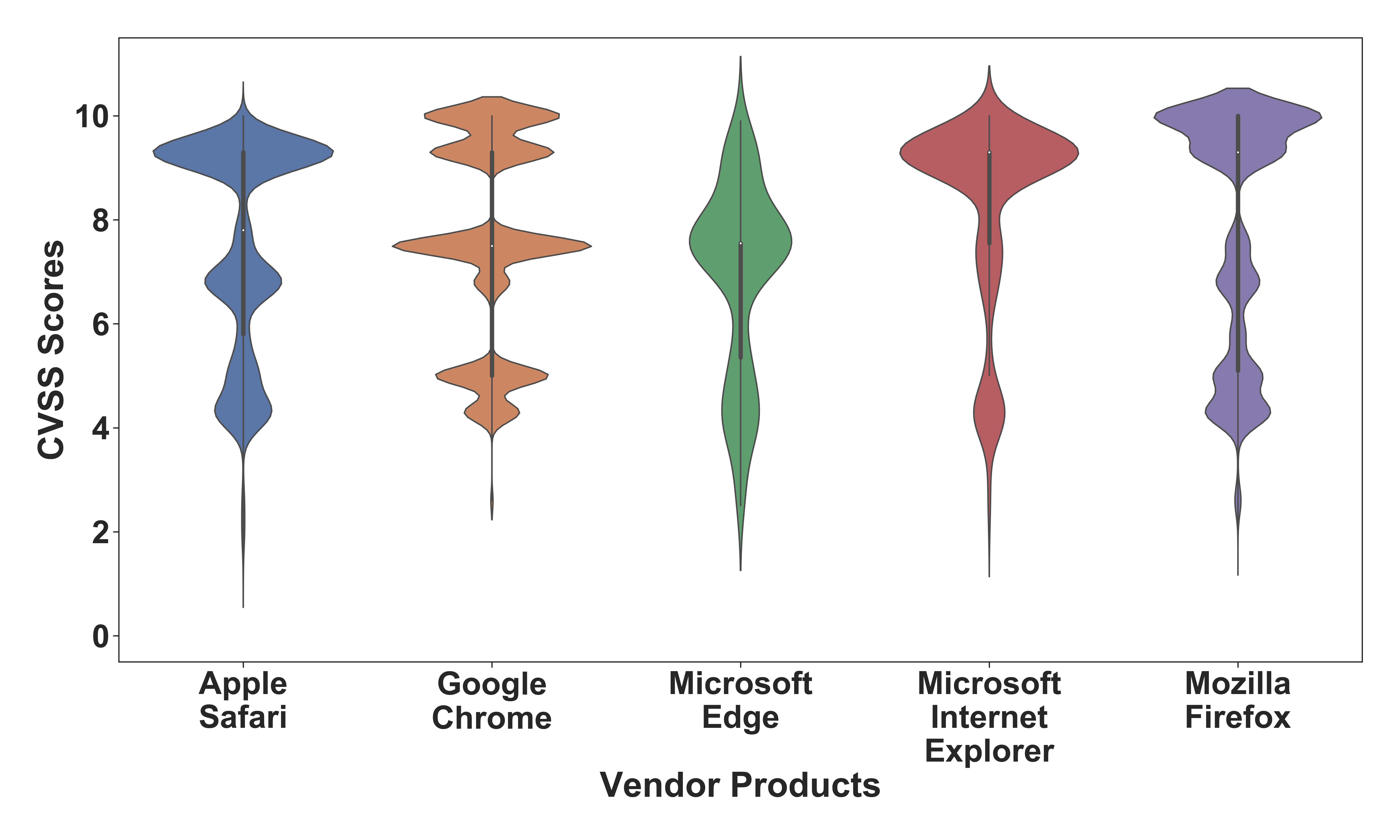}
  \caption{Distribution of severity scores for vendor browsers using all \VULNERABILITIES linked to all \config versions.}
  \label{fig:product-browser-all-cvss-violin}
\end{figure}

\begin{figure}[!bt]
  \centering
  \includegraphics[width=0.49\textwidth]{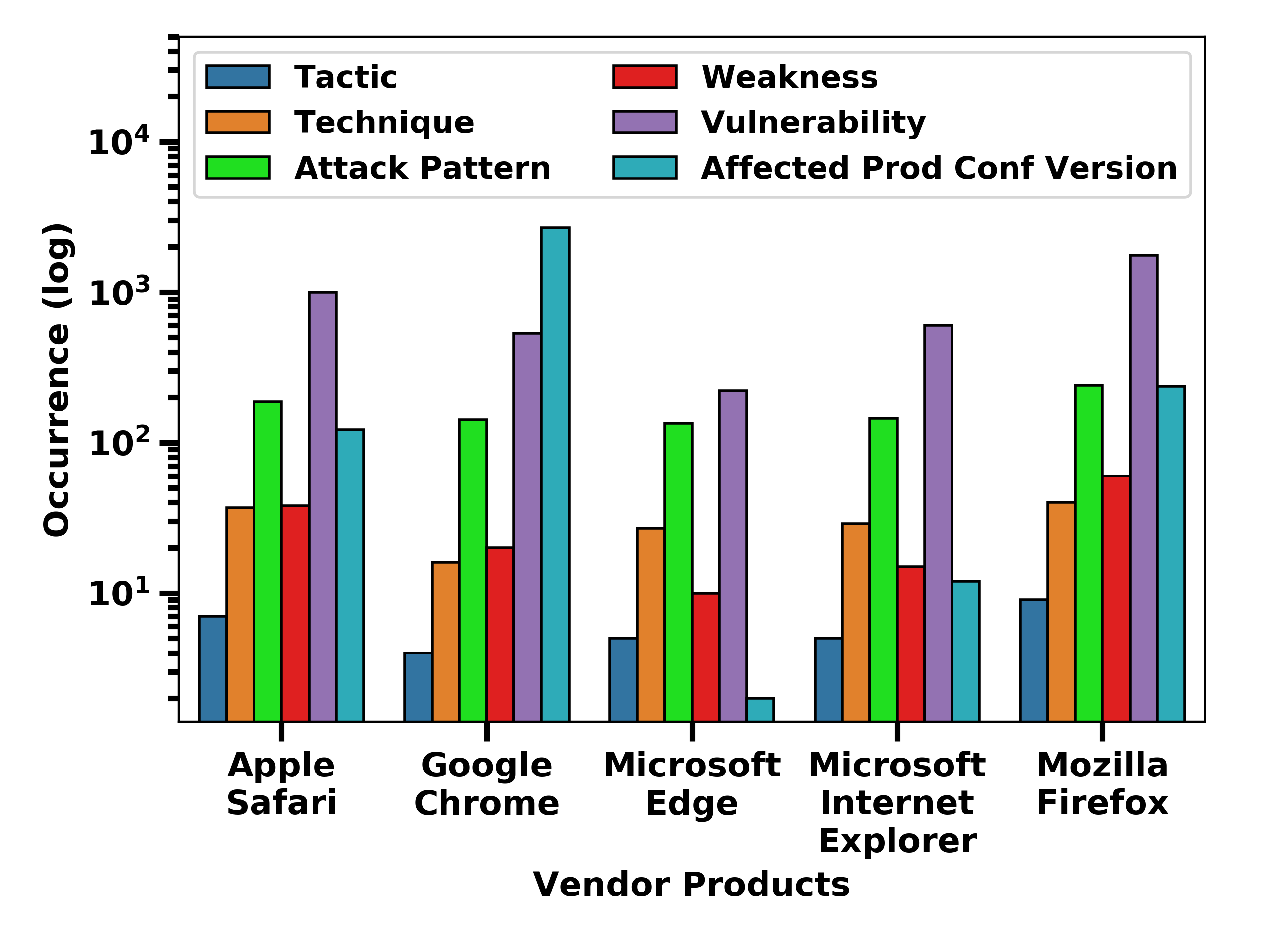}
  \caption{Number of unique data types for vendor browsers using all \config versions.}
  \label{fig:product-browser-bar}
\end{figure}

\begin{figure}[!bt]
  \centering
  \includegraphics[width=0.49\textwidth]{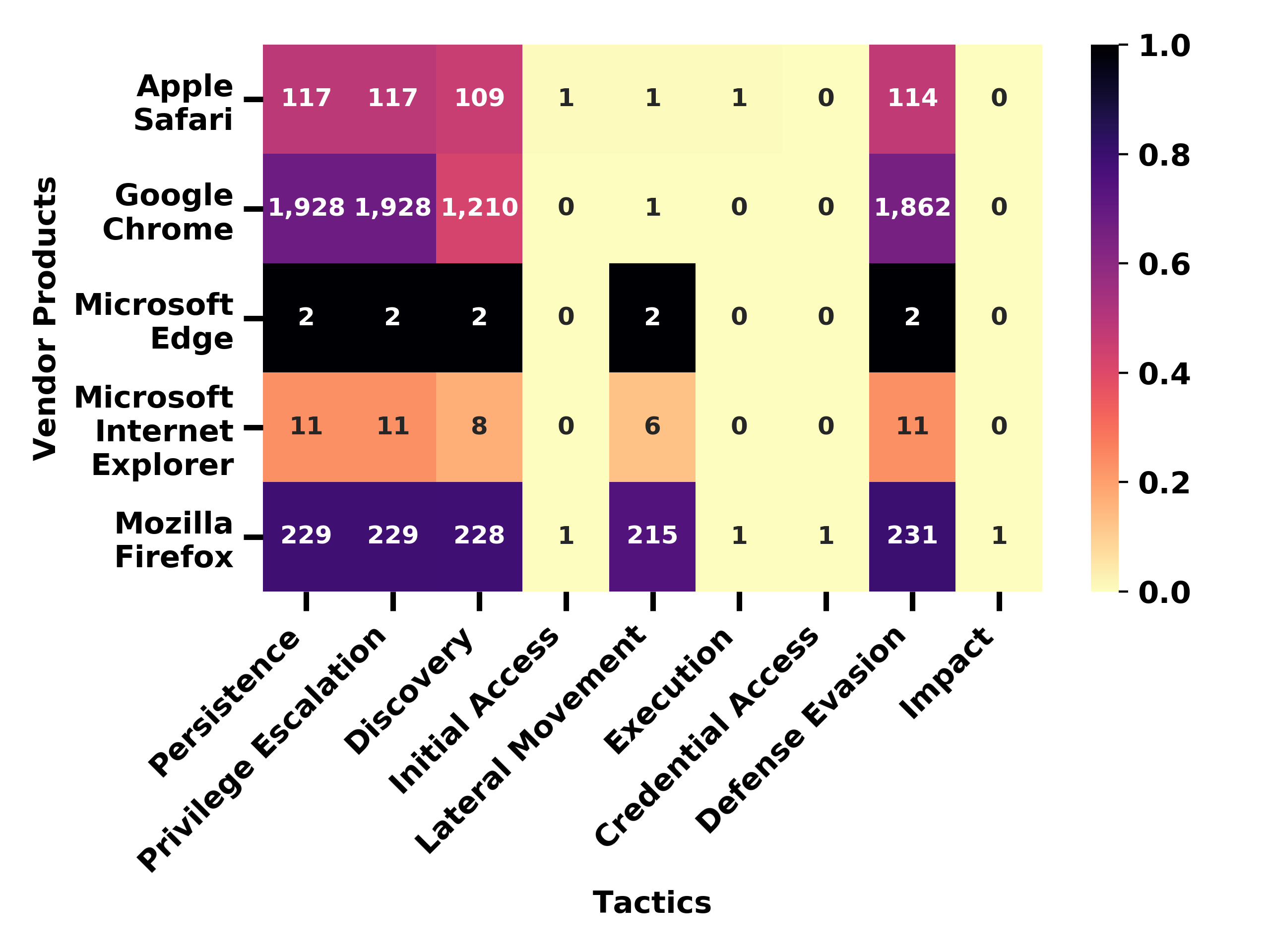}
  \caption{Heatmap showing number of product versions by vendor product that are exposed to nine \TACTICS. Colorbar represents the number of product versions exposed to the \TACTIC divided by the total number of product versions of the vendor product. Three \TACTICS are not shown because they did not expose any of these vendor products. Some product versions are exposed to multiple \TACTICS.}
  \label{fig:product-browser-heatmap}
\end{figure}

To explore this question, we chose a set of 5 web browsers: Apple Safari, Google Chrome, Mozilla Firefox, Microsoft Edge and Explorer.  Note that Edge is a Chromium based browser that is developed most recently,  with security in mind.  We then examined their security scores across all the occasions they were referenced as an affected product configuration of a \VULNERABILITY (CVE).  This examination included any version of the browser.   The distributions of each browser's severity scores is shown in Figure~\ref{fig:product-browser-all-cvss-violin}. These violin plots can be cross-referenced to the number of entries per source per Figure~\ref{fig:product-browser-bar}.  All browsers are identified in the same rough number of \TACTICS, \TECHNIQS, and \ATTACKPATTERN{s}.  We observed Edge to be referenced in significantly fewer \configs over all versions than any other browser. However, it had relatively similar counts of entries in the other sources, when compared to Explorer, another Microsoft browser. Figure~\ref{fig:product-browser-heatmap} shows how many tactics each browser is under threat from, for all its versions. We see that Firefox is exposed to the most tactics, and approximately 80\% of its product versions are exposed to the same 5 \TACTICS.

\subsection{US-CERT Alerts}\label{sec:alerts}
\subsubsection{Top 10 Routinely Exploited \VULNERABILITIES}
\label{sec:top10CVEs}

\begin{table}[bt]
  \scriptsize
  \centering
  \caption{Most frequent exploited vulnerabilities 2016-2019~\cite{top10CVE} analyzed in \bron. \APC is \config}
  \label{tab:topcve}
  \begin{tabular}{lllllll}
\textbf{\VULNERABILITY} & \textbf{CVSS} & \textbf{\#\TACTICS} & \textbf{\#\TECHNIQS} &
\textbf{\#\AttackPatterns} & \textbf{\#\WEAKNESS} & \textbf{\#\APC}\\
\hline
{CVE-2017-11882} & {8.55} & {0} & {0} & {12} & {1} & {4}\\
{CVE-2017-0199} & {8.55} & {0} & {0} & {0} & {0} & {9}\\
{CVE-2017-5638} & {10.0} & {3} & {5} & {50} & {1} & {53}\\
{CVE-2012-0158} & {9.3} & {0} & {0} & {3} & {1} & {29}\\
{CVE-2019-0604} & {8.65} & {3} & {5} & {50} & {1} & {4}\\
{CVE-2017-0143} & {8.7} & {0} & {0} & {0} & {0} & {0}\\
{CVE-2018-4878} & {8.65} & {0} & {0} & {0} & {1} & {3}\\
{CVE-2017-8759} & {8.55} & {3} & {5} & {50} & {1} & {8}\\
{CVE-2015-1641} & {9.3} & {0} & {0} & {0} & {1} & {11}\\
{CVE-2018-7600} & {8.65} & {3} & {5} & {50} & {1} & {4}\\
\end{tabular}
\end{table}

The Cybersecurity and Infrastructure Security Agency (CISA), the Federal Bureau of Investigation (FBI), and the broader U.S. Government routinely issue technical guidance to advise IT security professionals of the most commonly known \VULNERABILITIES exploited by foreign cyber actors. See ~\cite{top10CVE} and column one of Table~\ref{tab:topcve}  for an example. Information about each \VULNERABILITY is concrete and provided to encourage patching. For example, it includes \CONFIG{s} and names the exploit observed to have been used in actual attacks. 

An alert can be more helpful if abstract knowledge about it can be
collected.  Cyber hunting and security measures can be enhanced given
some understanding of why and how a foreign actor may be targeting the
network.  This can make risk decision-making more informed.

Some threat hunting questions for CVEs of
such alerts are: what \TACTIC{s} and \TECHNIQ{s} drive them and what
\ATTACKPATTERN{s} feature the abstract \WEAKNESS{es} they instantiate?
To understand the actual issues that threaten today’s systems, what
\WEAKNESS{es} are identifiable as common and high impact?  Are there
commonalities among the listed CVEs?

To answer these questions we use \bron to trace each CVE to its  \WEAKNESS{es}. It traces from these \WEAKNESS{es} to potential \ATTACKPATTERN{s}. Finally, it traces from the  \TECHNIQS  of an  \ATTACKPATTERN to the potential \TACTIC.  Table~\ref{tab:topcve} shows results for the
Top 10 Most Exploited Vulnerabilities 2016-2019 Alert~\cite{top10CVE}.
Note, CVE-2017-0143 does not list any connections to \WEAKNESS{es} or
\CONFIGS in the \bron version we used.
The top (most commonly) exploited CVE is \texttt{CVE-2017-11882}. It has
4 different \CONFIG{s} that all are symptomatic of the same \WEAKNESS:
\texttt{CWE-119, Improper Restriction of Operations within the Bounds of a Memory Buffer}.
This \WEAKNESS is a feature of 12 different \ATTACKPATTERN{s} though none
of these \ATTACKPATTERN{s} identify a \TECHNIQ.

The third ranked CVE, \texttt{CVE-2017-5638}, has the highest severity
score of 10.0. It lists 53 different \CONFIG{s} and instantiates only one
\WEAKNESS. However, this \WEAKNESS is a feature of 50 different
\ATTACKPATTERN{s}. Five different \TECHNIQS are involved in these
\ATTACKPATTERN{s}: \texttt{T1027, T1148, T1562.003, T1574.006, T1574.007}.
They are means of achieving three different \TACTICS:
\texttt{Defense Evasion, Persistence, Privilege Escalation}. This
information provides additional context for cyber hunting such as if any
\CONFIG{s} have not been patched and thus exposed to exploit.

This inquiry reveals disparities in the threat information related to the
Top 10 CVEs, and the path from CVE to \TACTIC is often cut short. Only 4
of the Top 10 CVEs are traceable to some \TACTIC and for each CVE, the
\TACTICS are Defense Evasion, Persistence, and Privilege Escalation. These
4 CVEs also share the same 5 \TECHNIQS, 50 \ATTACKPATTERNS, and 1 \WEAKNESS.
In total, given the information for these top 10 CVEs, 69 \CONFIG{s} are
under the threat of the three \TACTICS. Microsoft is the most listed
vendor among the \CONFIG{s} and most \PLATFORMS are Microsoft Office.



\subsubsection{CWE Top 25 Most Dangerous Software \WEAKNESS{es}}
\label{sec:cwe-top-25}

The \textit{2020 Common Weakness Enumeration (CWE) Top 25 Most Dangerous Software Weaknesses} list~\cite{topcwe} is compiled by considering the prevalence and severity of CVEs and their associated \WEAKNESS{es} (implemented by linking). These \WEAKNESS{es} highlight ``the most frequent and critical errors that can lead to serious vulnerabilities in software''~\cite{topcwe}. For example, an attacker can exploit the vulnerabilities to take control of a system, obtain sensitive information, or cause a denial-of-service.  The CWE Top 25 list is a resource that can provide insight into the most severe and current security weaknesses~\cite{topcwe}.

We use \bron to answer the following questions. What are the \TACTICS, \TECHNIQ{s} and \ATTACKPATTERNS linked to these \WEAKNESS{es}? What commonalities in these features are there among the Top 25?

\begin{table*}[tb]
\scriptsize
\centering
\caption{Top 25 CWE~\cite{topcwe}. \APC is \config.}
\label{tab:topcwe}
\begin{tabular}{lp{5cm}llp{1.5cm}llll}
\textbf{CWE ID} & \textbf{Name} & \textbf{\#\TACTICS} & \textbf{\#\TECHNIQS} &
\textbf{ \#\ATTACKPATTERNS} & \textbf{ \#\VULNERABILITIES} & \textbf{Sum CVSS} & \textbf{Ave CVSS} & \textbf{ \#\APC}\\
\hline

CWE-79
&
{Improper Neutralization of Input During Web Page Generation
('Cross-site Scripting')}
&
{0}
&
{0}
&
{6}
&
{12,629}
&
{57,890}
&
{4.58}
&
{43,013}
\\

CWE-787
&
{Out-of-bounds Write}
&
{0}
&
{0}
&
{0}

&
{1,159}
&
{8,667}
&
{7.48}
&
{1,499}
\\

CWE-20
&
{Improper Input Validation}
&
{1}
&
{3}
&
{51}
&
{7,820}
&
{49,995}
&
{6.39}
&
{44,399}
\\

CWE-125
&
{Out-of-bounds Read}
&
{0}
&
{0}
&
{2}
&
{2,172}
&
{13,549}
&
{6.24}
&
{2,029}
\\

CWE-119
&
{Improper Restriction of Operations within the Bounds of a Memory
Buffer}
&
{0}
&
{0}
&
{12}
&
{10,449}
&
{81,125}
&
{7.76}
&
{39,673}
\\

CWE-89
&
{Improper Neutralization of Special Elements used in an SQL Command
('SQL Injection')}
&
{0}
&
{0}
&
{6}
&
{5,477}
&
{41,309}
&
{7.54}
&
{14,685}
\\

CWE-200
&
{Exposure of Sensitive Information to an Unauthorized Actor}
&
{2}
&
{14}
&
{58}
&
{6,824}
&
{32,813}
&
{4.81}
&
{29,874}
\\

CWE-416
&
{Use After Free}
&
{0}
&
{0}
&
{0}
&
{1,187}
&
{8,742}
&
{7.37}
&
{1,636}
\\

CWE-352
&
{Cross-Site Request Forgery (CSRF)}
&
{0}
&
{0}
&
{4}
&
{2,377}
&
{16,319}
&
{6.87}
&
{11,646}
\\

CWE-78
&
{Improper Neutralization of Special Elements used in an OS Command ('OS
Command Injection')}
&
{0}
&
{0}
&
{5}
&
{724}
&
{6,131}
&
{8.47}
&
{3,325}
\\

CWE-190
&
{Integer Overflow or Wraparound}
&
{0}
&
{0}
&
{1}
&
{1,218}
&
{8,268}
&
{6.79}
&
{2,129}
\\

CWE-22
&
{Improper Limitation of a Pathname to a Restricted Directory ('Path
Traversal')}
&
{0}
&
{0}
&
{5}
&
{2,964}
&
{18,684}
&
{6.3}
&
{14,368}
\\

CWE-476
&
{NULL Pointer Dereference}
&
{0}
&
{0}
&
{0}
&
{1,019}
&
{5,994}
&
{5.88}
&
{2,342}
\\

CWE-287
&
{Improper Authentication}
&
{4}
&
{3}
&
{10}
&
{1,654}
&
{11,453}
&
{6.92}
&
{13,061}
\\

CWE-434
&
{Unrestricted Upload of File with Dangerous Type}
&
{0}
&
{0}
&
{1}
&
{562}
&
{4,315}
&
{7.68}
&
{1,370}
\\

CWE-732
&
{Incorrect Permission Assignment for Critical Resource}
&
{0}
&
{1}
&
{11}
&
{427}
&
{2,654}
&
{6.22}
&
{1,151}
\\

CWE-94
&
{Improper Control of Generation of Code ('Code Injection')}
&
{0}
&
{0}
&
{3}
&
{2,287}
&
{17,683}
&
{7.73}
&
{12,666}
\\

CWE-522
&
{Insufficiently Protected Credentials}
&
{5}
&
{15}
&
{9}
&
{277}
&
{1,548}
&
{5.59}
&
{923}
\\

CWE-611
&
{Improper Restriction of XML External Entity Reference}
&
{0}
&
{0}
&
{1}
&
{488}
&
{3,381}
&
{6.93}
&
{1,985}
\\

CWE-798
&
{Use of Hard-coded Credentials}
&
{0}
&
{0}
&
{2}
&
{244}
&
{1,919}
&
{7.87}
&
{543}
\\

CWE-502
&
{Deserialization of Untrusted Data}
&
{0}
&
{0}
&
{1}
&
{387}
&
{3,151}
&
{8.14}
&
{1,580}
\\

CWE-269
&
{Improper Privilege Management}
&
{0}
&
{0}
&
{3}
&
{1,095}
&
{7,421}
&
{6.78}
&
{3,770}
\\

CWE-400
&
{Uncontrolled Resource Consumption}
&
{0}
&
{0}
&
{3}
&
{728}
&
{4,459}
&
{6.13}
&
{4,303}
\\

CWE-306
&
{Missing Authentication for Critical Function}
&
{0}
&
{0}
&
{4}
&
{112}
&
{793}
&
{7.09}
&
{504}
\\

CWE-862
&
{Missing Authorization}
&
{0}
&
{0}
&
{0}
&
{190}
&
{1,162}
&
{6.12}
&
{527}
\\
\end{tabular}
\end{table*}


Our analysis of the Top 25 CWE~\cite{topcwe} is summarized in Table~\ref{tab:topcwe}. We
observe 4 of 25  \WEAKNESS{es} lack a presence in any \AttackPattern. We must recognize again the ambiguity around whether the absence is due to unfinished efforts or reflects real experience.
Reflecting diversity in threats that could target the \WEAKNESS{es}, there are 8 distinct \TACTICS associated with the Top 25.

In terms of commonalities, the most frequent \TACTICS associated with the Top 25 weakness{es} are \texttt{Defense Evasion, Privilege  escalation, Discovery}. Only 2 \TECHNIQ{s}, \texttt{T1148, T1562.003}  occur  more than once. The three most frequent \ATTACKPATTERNS are \texttt{Using Slashes in Alternate Encoding, Exploiting Trust in Client, Command Line  Execution through SQL Injection}.

Table~\ref{tab:topcwe} can be sorted on different columns, affording further ranked comparisons.
The three most frequent \VULNERABILITIES (i.e. top 3) occur 3 times each and are \texttt{CVE-2017-7778, CVE-2016-10164, CVE-2016-7163}. The three most frequent \shortconfig(s) are 3 different linux versions occurring 23, 24, 25 times respectively.   We also analyzed the \WEAKNESS text descriptions with a frequency analysis of unigrams and bigrams. \texttt{Buffer Overflow} emerged as a common \ATTACKPATTERN bigram.

We note again that not all weakness are linked with the same frequency to \ATTACKPATTERN{s}, \TECHNIQS and \TACTICS.  The ambiguity of this finding is also noted again: is absent data due to curation or lack of demonstrated evidence?  This prompts a second line of inquiry: an inventory of \bron in Section~\ref{sec:RelationalAnalysis}.

\section{Analyzing \bron's Sources}
\label{sec:RelationalAnalysis}
\label{para:aggregate counts and connections}

Many members of the security community rely upon the sources
amalgamated within \bron.  Evaluating how completely and accurately
these sources provide a view of a current threat landscape is easier
to attempt with \bron.  We can start an inventory of the sources and
their inter-connections by first analyzing the breadth and depth of
the entries and relations within \bron's graph. This involves counting
\bron's nodes and edges by type using different filters that offer
different resolutions. We proceed to answer some research questions (RQ).

In Section~\ref{para:aggregateRelations} we look at \bron to explore
how comprehensive (connected) the entries are. In
Section~\ref{para:VulnEdges} we investigate how the data has changed
over time. Finally, Section~\ref{sec:risks} explores the severity scores of the
publicly disclosed vulnerabilities.

\subsection{Aggregate Entry and Relational Analysis}\label{para:aggregateRelations}

\ResearchQ How comprehensively connected are the entries in \bron, i.e. sources of public data?

\begin{figure}[!tb]
  \centering
  \scalebox{.5}{\begin{tikzpicture}[
tactic/.style={circle, draw=blue, ultra thick, minimum size = 2.28+7 mm, inner sep = 2pt},
techniq1/.style={circle, draw=orange, ultra thick, minimum size = 1+7mm, inner sep = 2pt},
techniq2/.style={circle, draw=orange, ultra thick, minimum size = 5.79+7 mm, inner sep = 2pt},
techniq3/.style={circle, draw=orange, ultra thick, minimum size = 4.12+7 mm, inner sep = 2pt},
capec1/.style={circle, draw=green, ultra thick, minimum size = 5.04+7 mm, inner sep = 2pt},
capec2/.style={circle, draw=green, ultra thick, minimum size = 2.41+7 mm, inner sep = 2pt},
capec3/.style={circle, draw=green, ultra thick, minimum size = 3.73+7 mm, inner sep = 2pt},
capec4/.style={circle, draw=green, ultra thick, minimum size = 6.87+7 mm, inner sep = 2pt},
cwe1/.style={circle, draw=red, ultra thick, minimum size = 5.21+7 mm, inner sep = 2pt},
cwe2/.style={circle, draw=red, ultra thick, minimum size = 4.64+7 mm, inner sep = 2pt},
cwe3/.style={circle, draw=red, ultra thick, minimum size = 4+7 mm, inner sep = 2pt},
cve1/.style={circle, draw=violet, ultra thick, minimum size = 27.31},
cve2/.style={circle, draw=violet, ultra thick, minimum size = 28.8},
cve3/.style={circle, draw=violet, ultra thick, minimum size = 29.43},
cve4/.style={circle, draw=violet, ultra thick, minimum size = 12.26},
cpe/.style={circle, draw=cyan, ultra thick, minimum size = 25.68, inner sep = 2pt},
]


\node[minimum size = 20 mm, align=center] (tactic-type) at (-2.25,-1) {\huge \textbf{\TACTIC}};
\node[tactic] (tactic1) at (8,-1) {\huge \textbf{12}};

\node[minimum size = 20 mm, align=center] (technique-type) at (-1.6,-3) { \huge \textbf{\TECHNIQ}};
\node[techniq1] (technique1) at (5.5,-3) {\huge \textbf{1}};
\node[techniq2] (technique2) at (8,-3) {\huge \textbf{195}};
\node[techniq3] (techniq3) at (10.5,-3) {\huge \textbf{70}};

\draw[<->, very thick] (tactic1) -- (technique2);
\draw[<->, very thick] (tactic1) -- (techniq3);

\node[minimum size = 20 mm] (capec-type) at (-0.9,-5) {\huge \textbf{\ATTACKPATTERN}};
\node[capec1] (capec1) at (5.5,-5) {\huge \textbf{128}};
\node[capec2] (capec2) at (8,-5) {\huge \textbf{14}};
\node[capec3] (capec3) at (10.5,-5) {\huge \textbf{52}};
\node[capec4] (capec4) at (13, -5) {\huge \textbf{325}};

\draw[<->, very thick] (techniq3) -- (capec2);
\draw[<->, very thick] (techniq3) -- (capec3);

\node[minimum size = 20 mm] (cwe-type) at (-1.7,-7) {\huge \textbf{\WEAKNESS}};
\node[cwe1] (cwe1) at (8,-7) {\huge \textbf{142}};
\node[cwe2] (cwe2) at (10.5, -7) {\huge \textbf{100}};
\node[cwe3] (cwe3) at (13, -7) {\huge \textbf{72}};

\draw[<->, very thick] (capec3) -- (cwe2);
\draw[<->, very thick] (capec3) -- (cwe3);
\draw[<->, very thick] (capec3) -- (cwe1);
\draw[<->, very thick] (capec4) -- (cwe1);
\draw[<->, very thick] (capec4) -- (cwe3);

\node[minimum size = 20 mm] (cve-type) at (-1.2,-9.75) {\huge \textbf{\VULNERABILITY}};
\node[cve1] (cve1) at (5.5, -9.75) {\huge \textbf{20,383}};
\node[cve3] (cve3) at (8, -9.75) {\huge \textbf{25,496}};
\node[cve2] (cve2) at (10.5, -9.75) {\huge \textbf{23,980}};
\node[cve4] (cve4) at (13, -9.75) {\huge \textbf{1,856}};

\draw[<->, very thick] (cwe3) -- (cve2);
\draw[<->, very thick] (cwe2) -- (cve2);
\draw[<->, very thick] (cwe3) -- (cve3);
\draw[<->, very thick] (cwe2) -- (cve3);

\node[minimum size = 20 mm, align=left] (cpe-type) at (-0.5,-12.75) {\huge \textbf{\textit{Affected}} \huge \textbf{\textit{Product}} \\[1.5mm] \huge \textbf{\textit{Configuration}}};
\node[cpe] (cpe1) at (8, -12.75) {\huge \textbf{16,941}};

\draw[<->, very thick] (cve4) -- (cpe1);
\draw[<->, very thick] (cve2) -- (cpe1);

\end{tikzpicture}}
  \caption{Visualization of \bron with \VULNERABILITIES from only
    2015-2020 (subset of all \VULNERABILITIES in \bron) and only the latest version of \configs.}
  \label{fig:meta_results}
\end{figure}
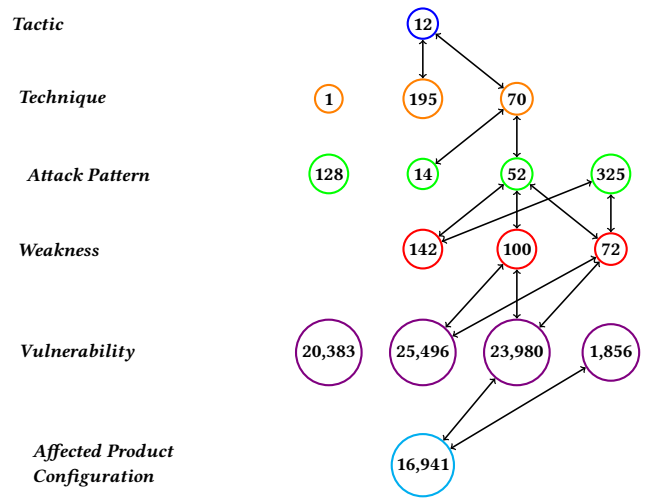

We start by aggregating the number of entries by information source and investigating the connectivity between them. By source, we visualize how many entries  are connected in Figure~\ref{fig:meta_results}.
\label{para:floaters}
We observe what we call ``\floaters'', i.e. entries that are isolated or ``orphaned''.
Only a small fraction of \TECHNIQ{s} are \floaters (1 of $266$).
By definition, no \WEAKNESS{es} or \configs are \floaters.
There are however two obvious sets of \floaters.  Of the $71,715$ \VULNERABILITIES, $28\%$ are floating entries.  Of the $519$ \ATTACKPATTERNS, $25\%$ are \floaters.  This draws attention to a level of incompletion within \bron. However, \bron's sources are constantly being updated and that \TECHNIQS are a more recent source of information.  A valid question is whether it will ever be possible to keep up.

\label{para:supernodes}
Figure~\ref{fig:meta_results} also indicate \supernodes which equate to specific entries that are the destination and/or source of many relational cross-references, i.e. nodes with many edges.
 We expect \supernodes given that there are many more \VULNERABILITIES and \configs than other entry types. These naturally will be more densely inter-connected.
%
We find that many \configs are different versions of the same software product.
Depending on how long a \VULNERABILITY has been latent, many historical versions of the software may be affected when it is identified.
For example, a vulnerability in Windows 10 may extend to all prior
versions of Windows.
\bron presently compiles \VULNERABILITY data dating back to 1999 so this could explain \supernodes. We therefore filtered out \configs that were not the most recent version of a software product and examined how this changed the quantities of relational connections  between \VULNERABILITIES and \configs.
The number of \configs decreased approximately $80\%$, from $219,767$ to  $42,061$ and the number of links of both \configs and \VULNERABILITIES also decreased significantly.

\label{para:Source-Centered Relationships}
\label{para:TACTIC edges}
\subsubsection{Data Source Analysis}

\ResearchQ How extensively are each of the different data sources in \bron connected?

\begin{figure*}[tb]
\centering
  \begin{subfigure}{0.32\textwidth}
  \centering
  \includegraphics[width=\textwidth]{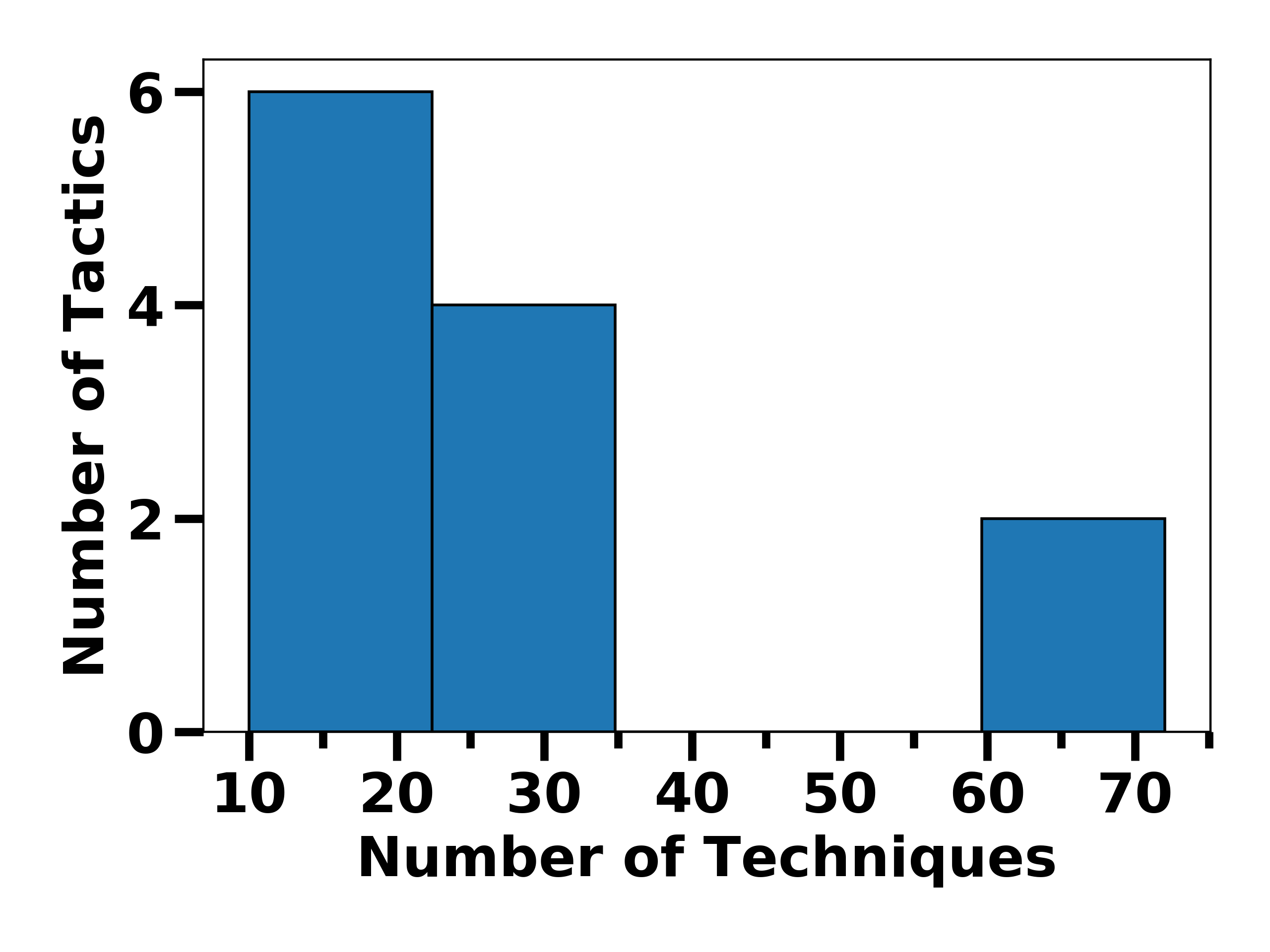}
\caption{}
  \label{fig:tactic_edges}
\end{subfigure}
  \begin{subfigure}{0.32\textwidth}
  \centering
  \includegraphics[width=\textwidth]{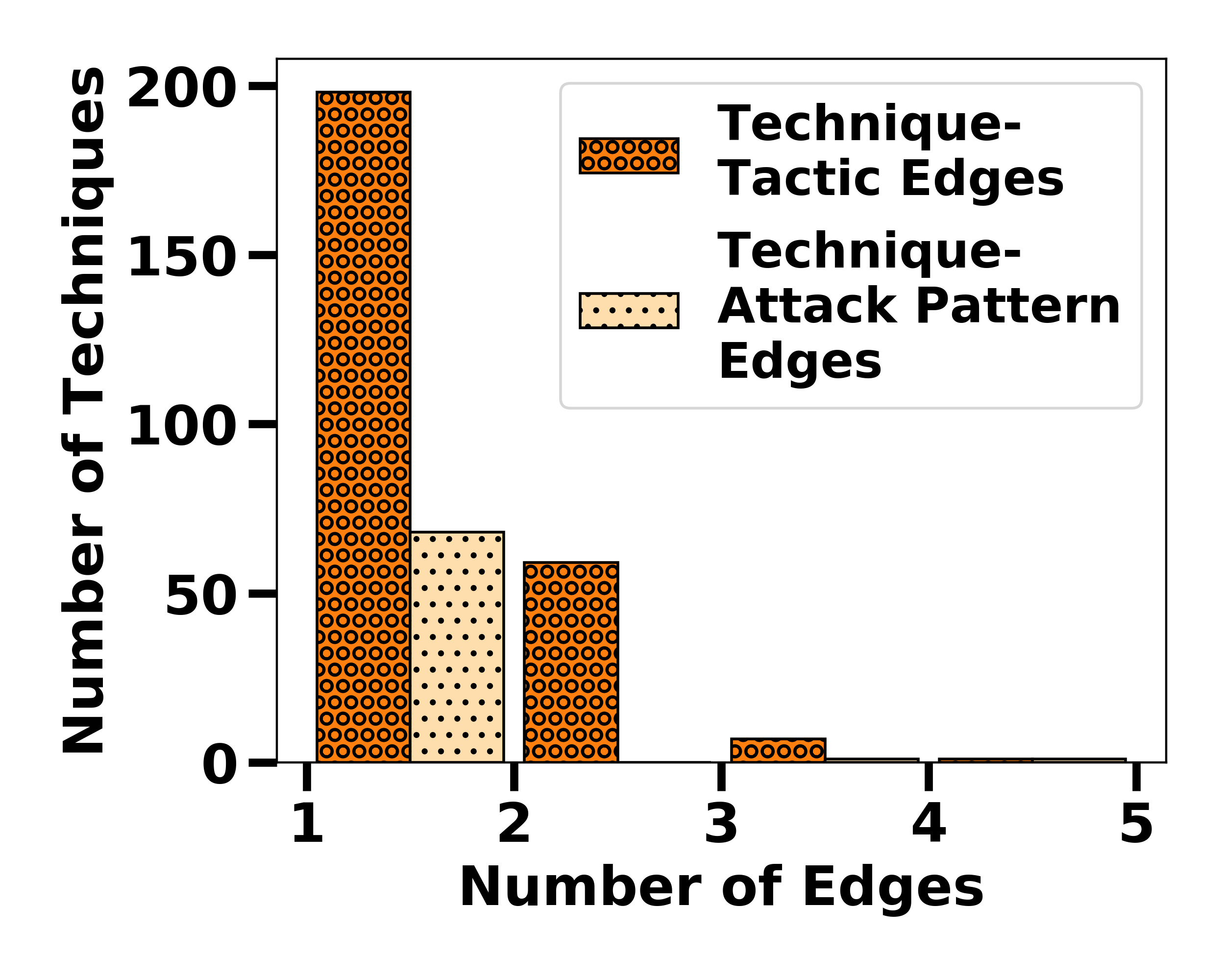}
\caption{}
  \label{fig:technique_edges}
\end{subfigure}
  \begin{subfigure}{0.32\textwidth}
  \centering
  \includegraphics[width=\textwidth]{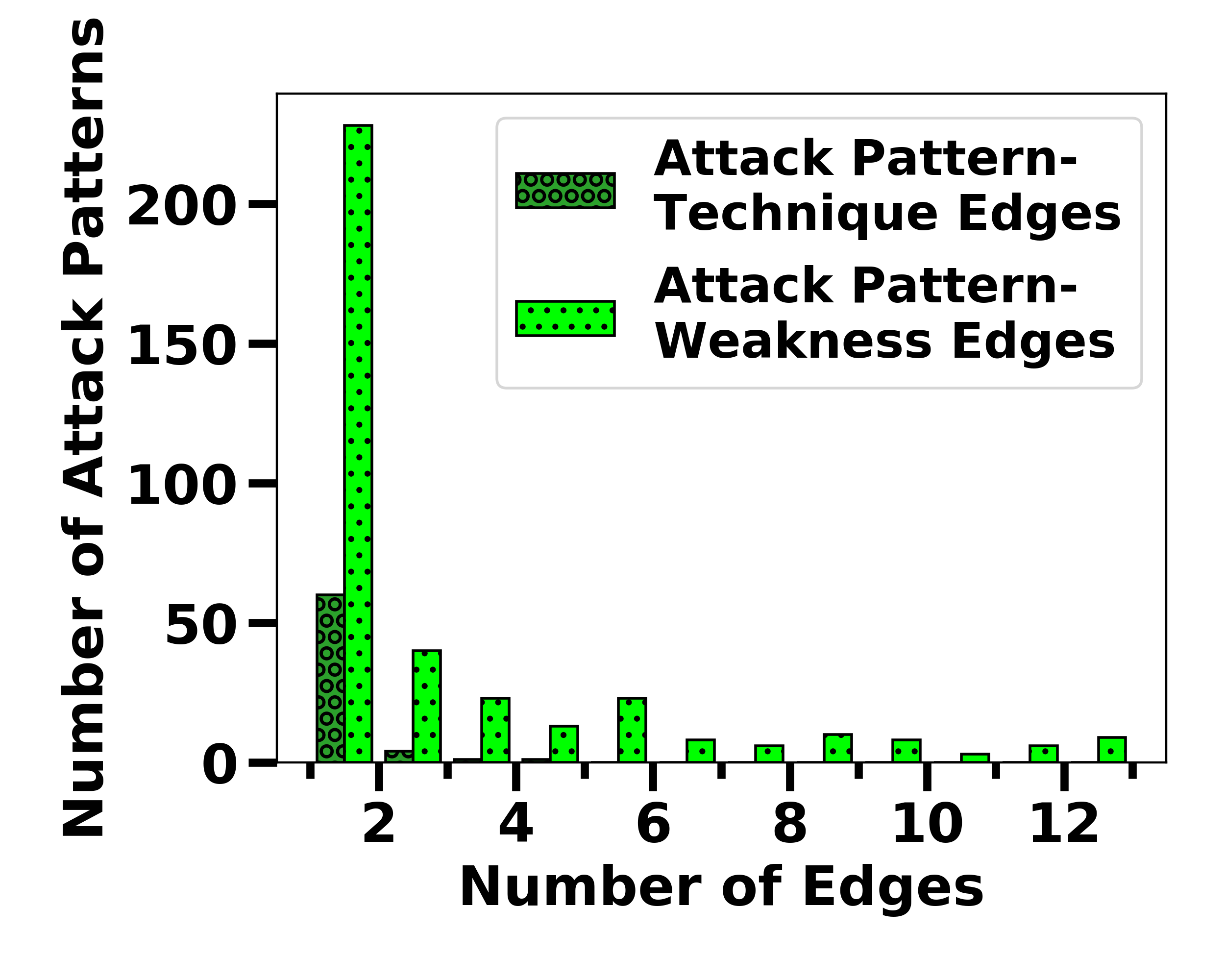}
\caption{}
  \label{fig:capec_edges}
\end{subfigure}
  \caption{Plots showing relational linkage statistics for  (a) \textbf{\TACTICS}, (b) \textbf{\TECHNIQS}, and (c)  \textbf{\ATTACKPATTERNS}.
}
  \label{fig:edges}
\end{figure*}

Starting with \TACTICS in Figure~\ref{fig:tactic_edges}, we find that all of the \TACTICS are linked to at least 10 \TECHNIQS.
Recall ATT\&CK is a matrix of \TECHNIQS, organized by a column per \TACTIC, and a link implies that a \TECHNIQ serves the tactical objective named by the \TACTIC.
The number of \TECHNIQS serving  each \TACTIC varies. Several \TACTICS are served by a number of \TECHNIQS.
However, the magnitude of the difference between the \TACTIC with the least and most links is $62$ (while the median number of relational links is 22.5).
The \TACTIC served by the most \TECHNIQS is \texttt{Defense Evasion} which is linked to 72 \TECHNIQS.

\label{para:TECHNIQUE Edges}
\TECHNIQS  can serve more than one \TACTIC, and we observe that most of the 266 \TECHNIQS serves at least one \TACTIC but very few serve multiple \TACTICS. The \TECHNIQ \texttt{Valid Accounts} serves two \TACTICS, and
six \TECHNIQS each serve three \TACTICS.
We can inventory \TECHNIQ and \ATTACKPATTERN relationships. The relationship associates the means of an attack, i.e. \TECHNIQS, to the attack itself. We find that
 74\% of the \TECHNIQS are not associated with any \ATTACKPATTERN. Each of the remaining \TECHNIQS, except for two, serve only one \ATTACKPATTERN.
The two exceptions are (1) the \texttt{Supply Chain Compromise} which serves three \ATTACKPATTERN{s} and (2) \texttt{Endpoint Denial of Service} which serves four \ATTACKPATTERN{s}.
From a \TECHNIQ analysis, an overall lack of relational information connecting the objectives of \TACTICS and the means of \TECHNIQS with \ATTACKPATTERN{s} becomes apparent. We will next look at \ATTACKPATTERN{s} to consider their relations to \TECHNIQS and, further, inquire into the \ATTACKPATTERNS that connect \TECHNIQS and \WEAKNESS{es}.

\label{para:ATTACKPATTERNS}
Pivoting to \ATTACKPATTERNS, we can recall that of the $519$ \ATTACKPATTERNS, $25\%$ are \floaters.  They are named attacks with no indication of how they could be accomplished or  what weakness in a system they would exploit.
We observe that only $10\%$ of \ATTACKPATTERNS span both \TECHNIQS and \WEAKNESS{es}.
Per Figure~\ref{fig:capec_edges}, most \ATTACKPATTERNS have only one relational link and
an \ATTACKPATTERN has at most $13$.
There are $3$ \ATTACKPATTERNS with $13$ relational links:
\begin{inparaenum}
\item \texttt{Using Leading 'Ghost' Character Sequences to Bypass Input
  Filters},
\item \texttt{Manipulating Web Input to File System Calls}, and
\item \texttt{Using Slashes in Alternate Encoding}.
\end{inparaenum}
All $13$ links of each of these \ATTACKPATTERNS  span from \ATTACKPATTERN to \WEAKNESS.
None of the links relate the \ATTACKPATTERN to a \TECHNIQ.
The \ATTACKPATTERN \texttt{Remote Services with Stolen Credentials} identifies the most
means of attack with 4 \TECHNIQ links. It also links to one \WEAKNESS: \texttt{CWE-522}~\texttt{Insufficiently Protected Credentials}.

\label{para:GapAttackPatterns}
One can think of an ideal  \ATTACKPATTERN as a bridge. It connects an attack to how it is executed and to where it is targeted. At this point in time, \bron users could benefit from more connected relational bridges. It is reasonable to expect that future efforts will more completely enumerate \ATTACKPATTERN{s} to assure a more relationally informative set of sources. However, given threat assessment technologies make use of these sources, this inventory-based analysis reveals possible blind spots or unfinished curation.  The ambiguity between the two possibilities is hard to resolve.

\label{para;Weakness Relations}

%

\begin{figure}[!tb]
  \centering
  \includegraphics[width=0.49\textwidth]{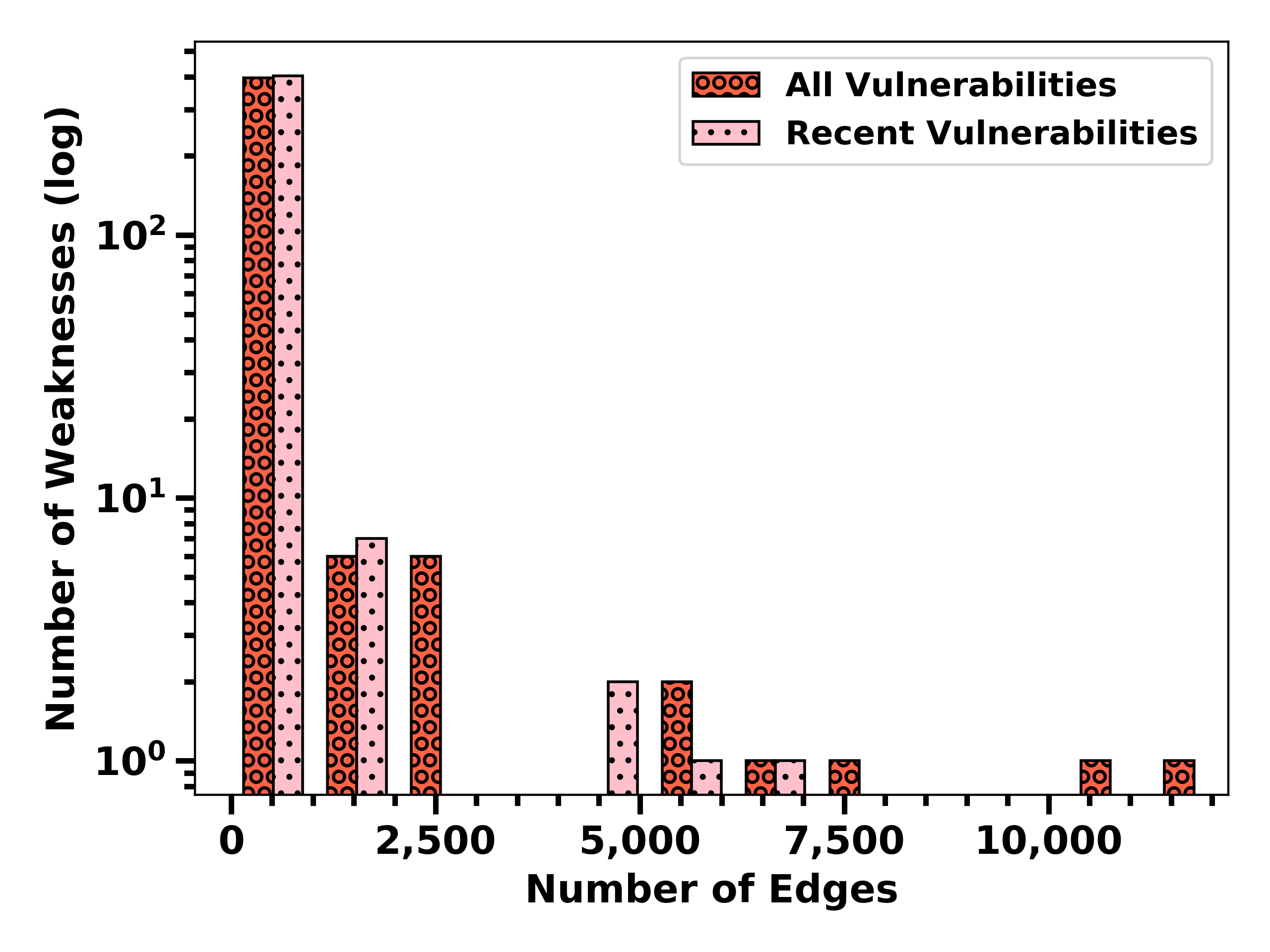}
  \caption{Relational linkage statistics for \WEAKNESS entries when all \VULNERABILITIES and only more recent \VULNERABILITIES are counted.
}
  \label{fig:cwe_edges}
\end{figure}

 Analyzing the number of links to and from each \WEAKNESS entry,
we observe a large comparative increase (orders of magnitude) in the range, see Figure~\ref{fig:cwe_edges}, versus the information sources we have just analyzed.  When, regardless of year, relationships to and from each \WEAKNESS entry  are counted, the range is the highest for all information sources at $12,291$. The facts that \VULNERABILITIES, to which \WEAKNESS{es} connect, have the
greatest quantity of entries and that  \WEAKNESS{es}  are general abstract concepts which explains this ranking.

Around $45\%$ of \WEAKNESS{es} are targets identified by \ATTACKPATTERNS. Does this imply that there exist no attacks  targeting $55\%$ of the \WEAKNESS{es}, or is the record lacking? The \WEAKNESS{es} with large number of relations are mostly connected to \VULNERABILITIES and are identified as the targets of attacks. The \WEAKNESS \texttt{Information Exposure} is a target of $57$  \ATTACKPATTERN{s} (the highest number among \WEAKNESS{es}) and it further details $6,624$ \VULNERABILITIES, i.e. more specific targets.
 The \WEAKNESS CWE-79,  \texttt{Improper Neutralization of Input During Web Page Generation}, \texttt{(Cross-}\texttt{Site Scripting)} connects to the most  \VULNERABILITIES at $1,292$. Six \ATTACKPATTERN{s} identify this \WEAKNESS as a target.
However, 44\% of these \VULNERABILITIES date before 2015.
When we focus only on \WEAKNESS{es} from 2015-2020, the range of number of \VULNERABILITIES decreases
significantly but is still quite substantial with several \supernodes. In the future, we can look at parent-child connections within \WEAKNESS{es} to better understand the extent of comprehensive linking.

\begin{figure}[!tb]
  \centering
  \includegraphics[width=0.49\textwidth]{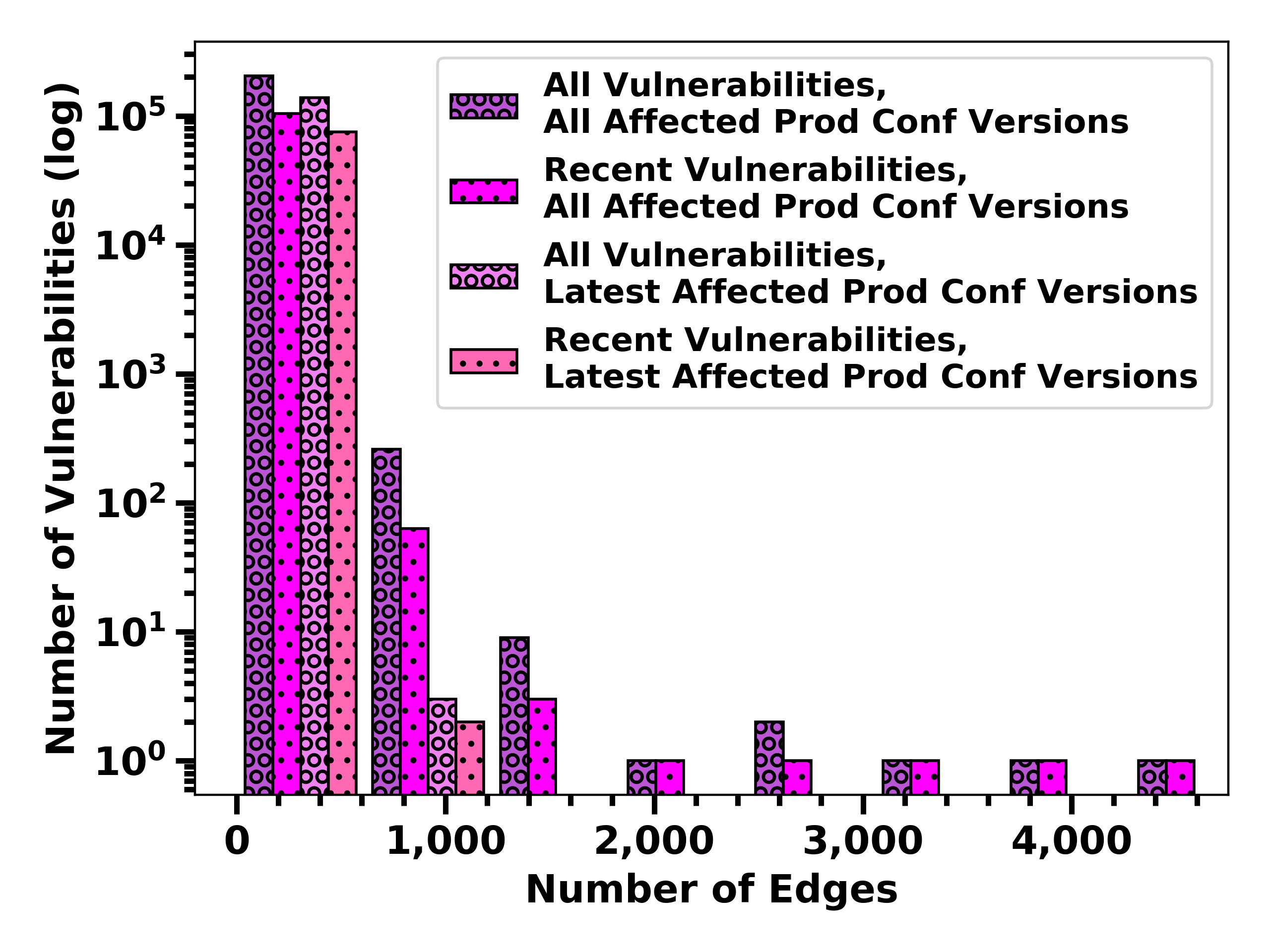}
  \caption{Count of \VULNERABILITY connections when all \VULNERABILITIES or more recent \VULNERABILITIES are used, and when all versions or only the latest version of \configs are referenced.
}\label{fig:cve_years}
\end{figure}

Moving to \VULNERABILITIES, most \VULNERABILITIES on average, regardless of
what years they were reported and how many \config versions are referenced, tend to connect
to 600 or fewer entries, either \WEAKNESS{es} or \configs. Figure~\ref{fig:cve_years} shows four combinations of connections: all or recent \VULNERABILITIES, and all or latest \configs.  The distribution when all \config versions are referenced is right-skewed with a long tail. The distribution when only the latest version of \configs is used is also right-skewed but has a shorter tail.    As would be expected when
removing earlier versions of a software application from consideration, many  \VULNERABILITY \floaters emerge and the number of \supernodes  decreases drastically.  When we filter down to only recent \VULNERABILITIES, while considering all
\config entries, we observe an increase in the number of \floater \VULNERABILITIES but we see a decrease when the number of \config entries decreases.

Severity scores are assigned by CVE Numbering Authorities (CNAs).  All 10 members of this vendor set are CNAs.
Although CVSS is an industry-standard, the calculated severity score within a vendor and among vendors is likely to vary. Variance can depend on the interpretation and details within vulnerability report. Additionally, analysts who calculate severity scores for vulnerabilities that do not contain sufficient details and data assume a worst-case scenario and assign a 10 to  \VULNERABILITIES with no information.\footnote{\url{https://nvd.nist.gov/vuln-metrics/cvss}}
As a result, some vendors may have \VULNERABILITIES with higher severity scores because of missing data.

The \VULNERABILITY with the most connections, when all versions of an \config are
considered, is \texttt{CVE-2016-1409}. This describes a bug in Cisco IOS that
creates a vulnerability to Denial of Service attacks. It has a severity score of
$6.25$ (on the CVSS  scale of $0$ to $10$). However, the \VULNERABILITY with the most connections changes when only the latest version of any  \config is considered. It is \texttt{CVE-2020-0551} which describes a vulnerability in an Intel product with a severity score of $3.75$.

In general, the large portion of floating \VULNERABILITIES indicates a
gap between the data informing its users of an \textit{abstract}
vulnerability. That is, what \textit{types or features} of products pose
risk versus informing them of an \textit{operational} vulnerability
that indicates \textit{specific products and versions} that are
affected and indicates what may need to be patched or updated. In
addition, Figure~\ref{fig:cve_years} shows difference of which CVEs to
select and which versions to use for analysis. It verifies what we
expected regarding version ``noise''. This can be useful when
considering versions of applications.

\label{para:AffectedConfigs}

\begin{figure}[!tb]
  \centering
  \includegraphics[width=0.49\textwidth]{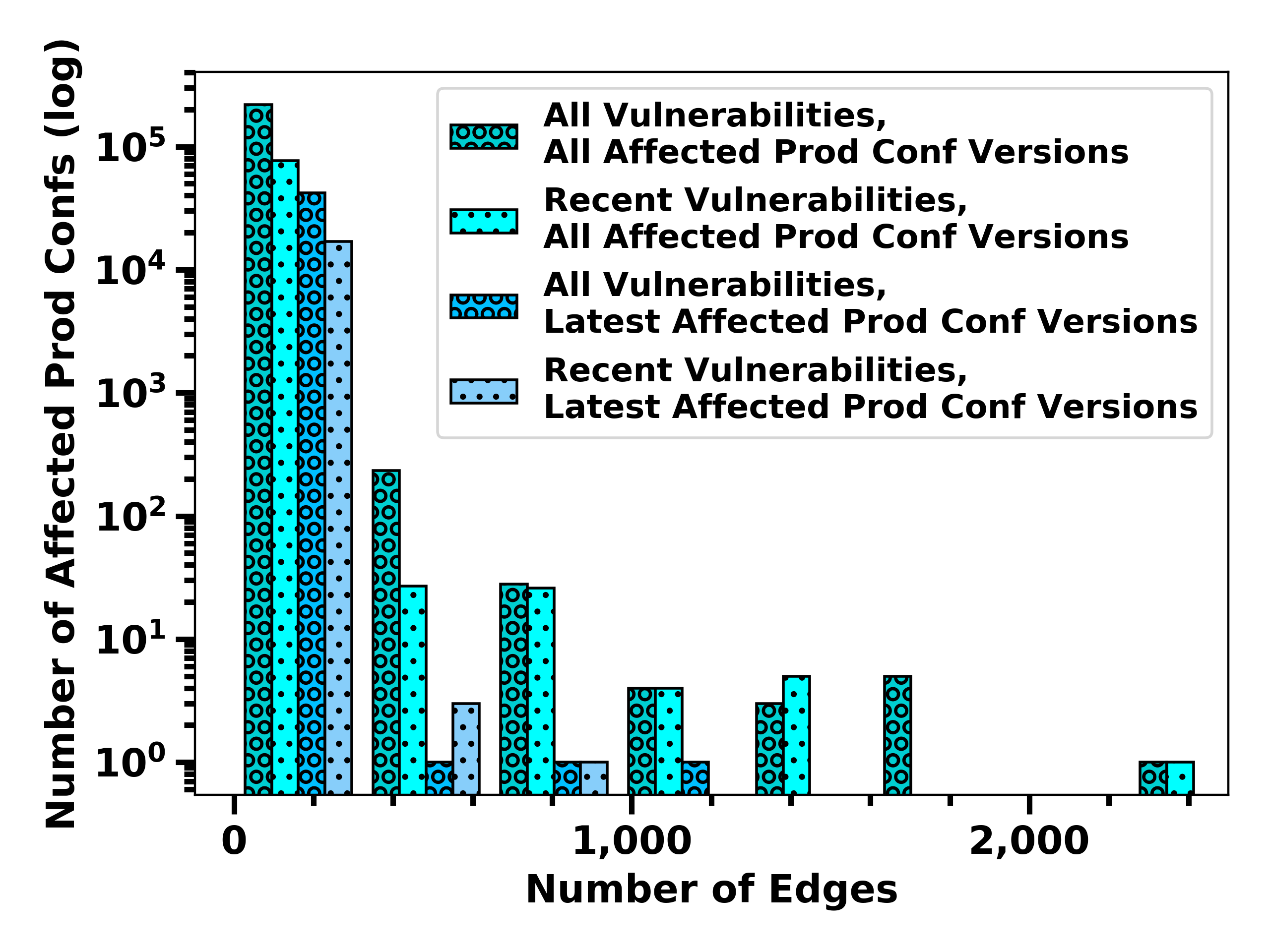}
  \caption{Count of \config connections when all \VULNERABILITIES or more recent \VULNERABILITIES are considered, and when all versions or only the latest version of a software product are counted.
}
  \label{fig:cpe_years}
\end{figure}

\config connection statistics are shown in Figure~\ref{fig:cpe_years}. The distribution when all  versions are referenced is right-skewed with a long tail. The distribution when only the latest version of \configs is referenced is right-skewed but has a shorter tail.  The vendor
and product that has the most connections spanning \VULNERABILITIES and its
versions is Linux Version 8 from Debian. However, when we
only consider one version of a \config, the \config with the most
connections is a Windows Server from 2012.
The Debian Linux 8 connections decrease because there are more recent versions of Debian Linux.
We see a similar pattern when
focusing on more recent \VULNERABILITY data. Debian Linux 8 is the most
commonly linked \config when all versions are considered, but a version of Windows 8 RT (which is discontinued) is the most commonly linked when only the latest version of a product is
considered. We can observe how the tail of the connection distribution shrinks as more recent data is in focus and multiple older versions are filtered, see
Figure~\ref{fig:cpe_years}.

We provide precise aggregate source and connective quantities in Table~\ref{tab:edge_counts}. We consider every version of a \config. The mean number of links from a \TACTIC to a \TECHNIQ is $28.4$ (stdev $18.9$), and the mean number of \WEAKNESS links is $8.16$ (stdev $46.70$).

\begin{table}[tb]
\caption{\bron source and relational quantities for non-\floaters. Median refers to the median number of relations in or out of an entry.
The rightmost column provides the lowest and highest counts of entries’ relations with the range in brackets.
}
\label{tab:edge_counts}
\small
\centering
\begin{tabular}{l|p{1cm}p{1.5cm}p{2.5cm}}
\textbf{Source} & \textbf{Total Entries} & \textbf{Median \#Links} & \textbf{Count Range } \\
\hline
\TACTIC & 12 & $22.5$ & 10-72 (62) \\
\TECHNIQ & 266 & 1 & 1-4 (3) \\
\ATTACKPATTERN & 519 & 1 & 1-13 (12) \\
\WEAKNESS & 314 & 4 & 1-12,292 (12,291) \\
\VULNERABILITY & 144,594 & 1 & 1-4,891 (4,890) \\
\shortconfig & 219,767 & 1 & 1-2,573 (2,572) \\
\end{tabular}
\end{table}

\subsection{Vulnerability paths}
\label{para:VulnEdges}

\ResearchQ Does the coverage of the threats become more comprehensive over time, i.e. have the paths starting from \VULNERABILITIES changed?

\begin{figure*}[tb]
\begin{subfigure}{0.48\textwidth}
  \centering
  \includegraphics[width=1.0\textwidth]{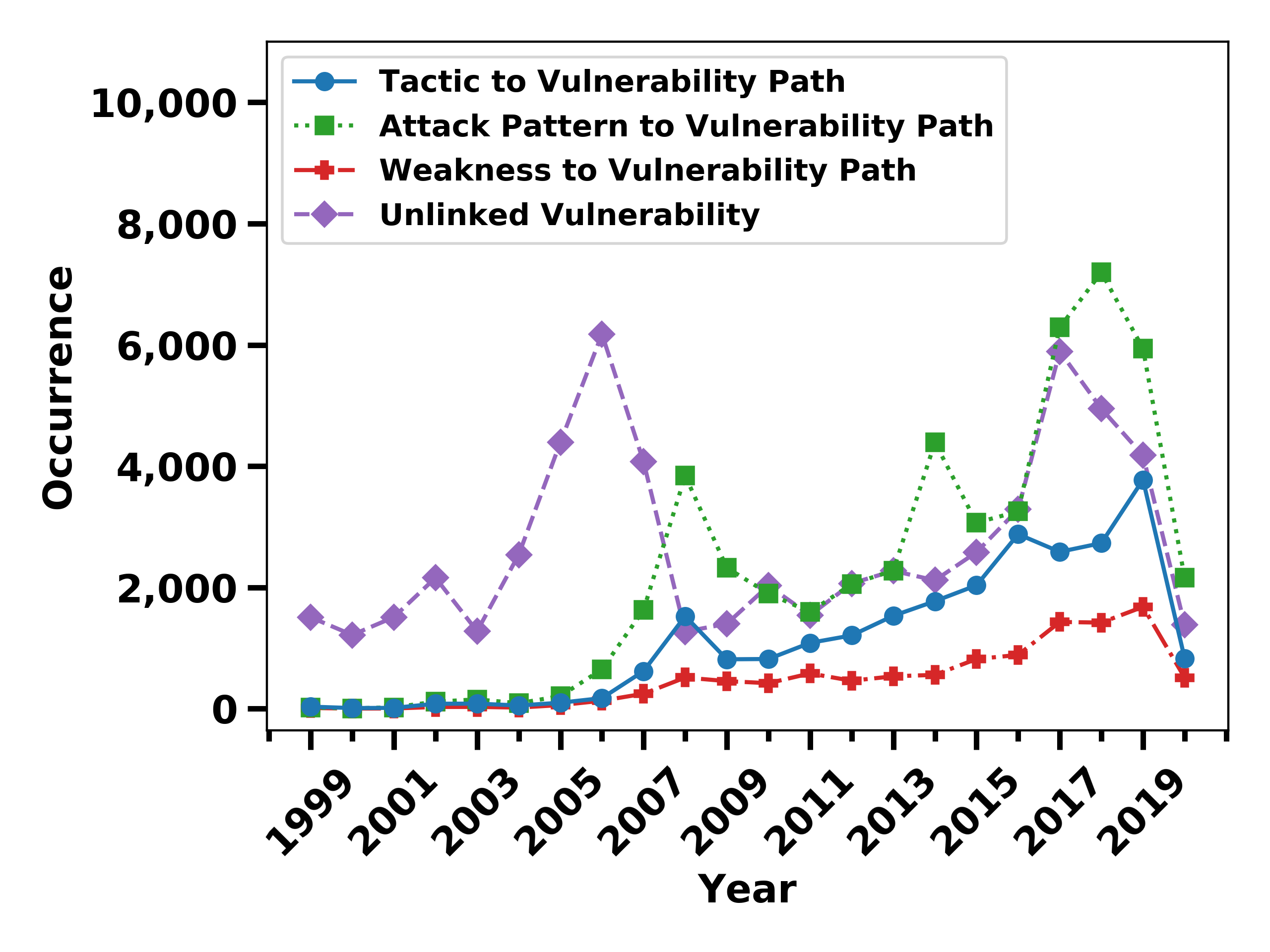}
  \caption{Number of \VULNERABILITIES reported by year.}
  \label{fig:cve_connect_to_data_types}
\end{subfigure}
\begin{subfigure}{0.48\textwidth}
  \centering
  \includegraphics[width=1.0\textwidth]{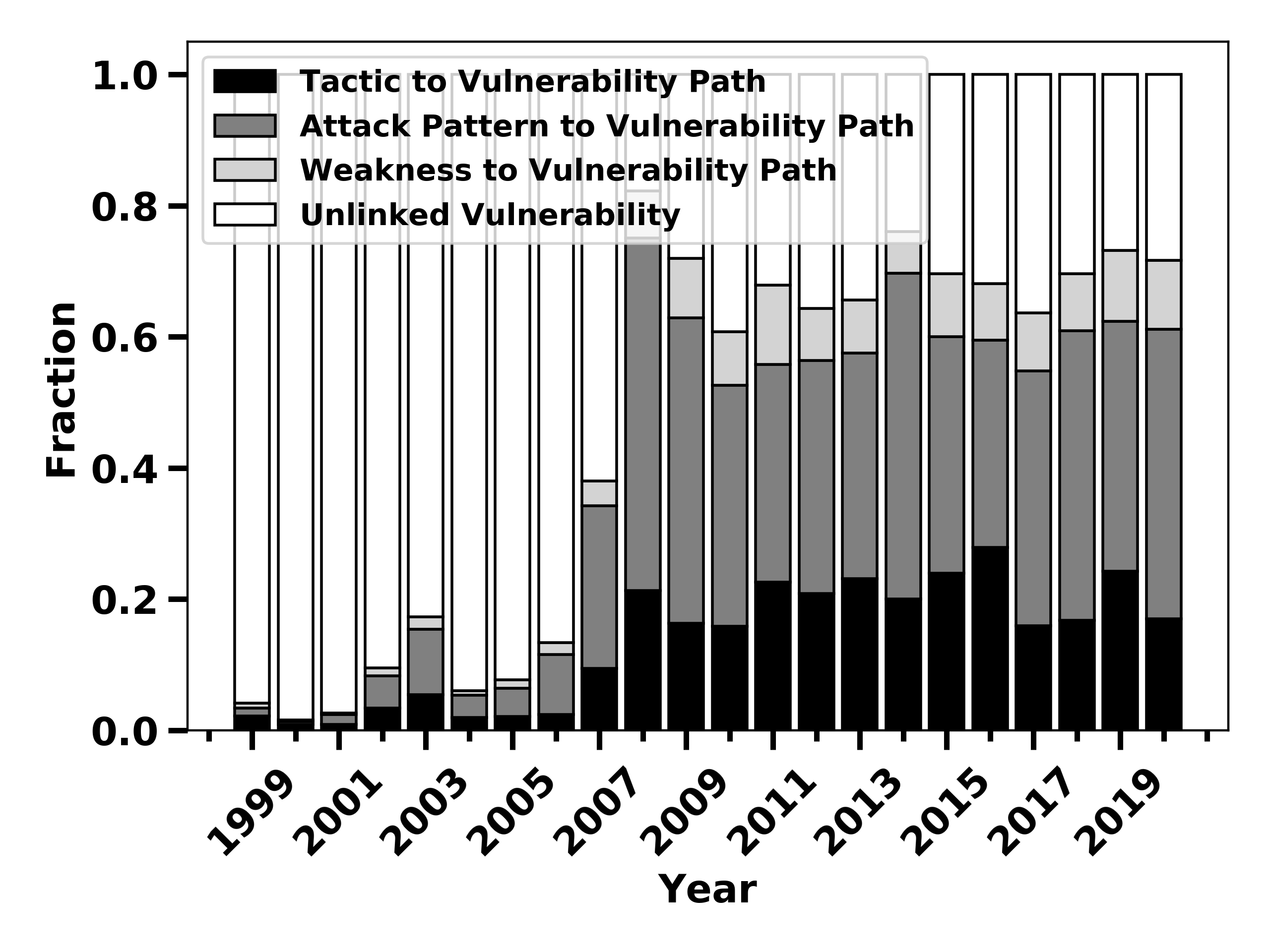}
  \caption{Percentage of \VULNERABILITIES reported by year.}
  \label{fig:cve_connect_to_data_types_pct}
\end{subfigure}
  \caption{Number and percentage of \VULNERABILITIES reported by year connected to a \TACTIC, \ATTACKPATTERN, or \WEAKNESS. Note 2020 has only $\approx 5$ months of data in these plots. In general, the number and percentage of \VULNERABILITIES connected to each data type is increasing over the years.}
  \label{fig:cve_connect_to_data_types_both}
\end{figure*}

We consider paths connecting through \VULNERABILITIES, calling them \VULNERABILITY paths.
Figure~\ref{fig:cve_connect_to_data_types} shows yearly trends of how many such paths start from, respectively, a \TACTIC, an \ATTACKPATTERN, or a \WEAKNESS, and end at a \VULNERABILITY.
It also shows how many \VULNERABILITIES are not connected to a \WEAKNESS. Another interesting aspect of \VULNERABILITY entries is how their quantity has changed over time. We examined this using Figure~\ref{fig:cve_connect_to_data_types}. Over time, more \VULNERABILITIES are reported.  This may be due to more products use and more bug reporting.
When we filter down to only recent \VULNERABILITIES, while considering all
\config entries, we observe an increase in the number of \floater \VULNERABILITIES but we see a decrease when the number of \config entries decreases. Figure~\ref{fig:cve_connect_to_data_types_pct} shows that the ratios of paths from \VULNERABILITIES to other data types have been relatively stable since 2008.

\subsection{\bron Severity Scores}\label{sec:risks}

\ResearchQ Are newly discovered \VULNERABILITIES becoming more severe? Are the CVSS scores uniformly distributed over time or value? Are the \VULNERABILITY \floaters  harmless?

\begin{figure}[!bt]
  \centering
  \includegraphics[width=0.49\textwidth]{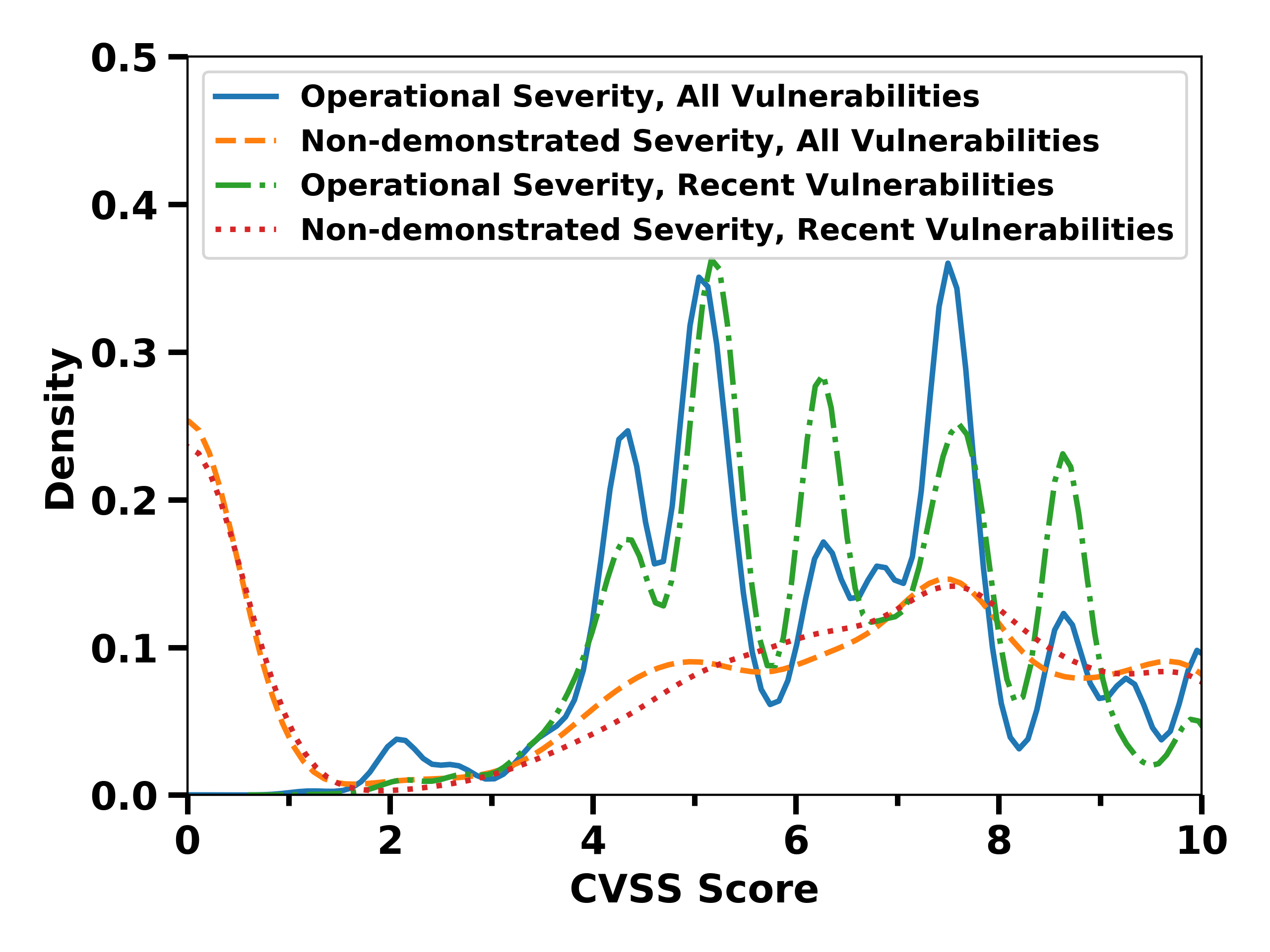}
  \caption{Severity scores when referencing all \VULNERABILITIES from 1999-2020 or only recent \VULNERABILITIES from 2015-2020. \Linked and \unlinked severity are shown.
}
\label{fig:risk}
\end{figure}

Because of the large number of \floaters, especially \VULNERABILITY \floaters, we
define  \bron's  ``\unlinked''  severity as the sum of CVSS severity scores assigned to \VULNERABILITIES which are not yet linked to any \configs i.e.  specific products and versions.
It contrasts with ``\linked'' severity which is the sum of severity scores assigned to \VULNERABILITIES which have at least one \config.
We define total severity as the sum of \unlinked and linked severity.
In general, \floater \VULNERABILITIES tend to have low severity scores (see Figure~\ref{fig:risk}).
This is regardless of whether all or most recent \config versions are considered.
  Nearly 25\% of \unlinked \VULNERABILITIES have severity scores of 0.
Recent \VULNERABILITIES  contribute more to \unlinked severity than those of previous years.
A possible explanation for this could be that it takes time (in the scale of years) after a \VULNERABILITY is documented to identify a specific \config. \cite{feutrill2018effect} discuss the release of only CVEs in more detail.

The number of \VULNERABILITIES with lower severity scores increases over the years, and most of these \VULNERABILITIES are not connected. For example, the percentage of \VULNERABILITIES that are not connected to  \WEAKNESS{es} decreased from 87\% in 2006 to 18\% in 2008, see Figure~\ref{fig:cve_connect_to_data_types_pct}, while the number of severity scores of 0 for those \VULNERABILITIES has increased between 2006 and 2008.  This correlation between the number of \floater  \VULNERABILITIES and the number of low severity scores suggests that  \VULNERABILITIES not linked to \WEAKNESS{es} tend to have lower severity scores. We observe a similar situation in that \VULNERABILITIES not connected to \configs tend to have lower severity scores as seen in Figure~\ref{fig:risk}. We observe a similar situation in that \VULNERABILITIES not connected to \configs tend to have lower severity scores, see Figure~\ref{fig:risk}.

The amount of \floater severity was unexpected when we tried to answer
questions about where the severity was, see Figure~\ref{fig:risk}. The length of
the tails for the different public threat data types are very
interesting. Does that imply categorization error (should have more
refined categories) or that some analysts spend more effort finding and writing out the vulnerabilities?

\label{para:end}  

\section{\bron useability}
\label{sec:bron-impl-perf}

\bron is intended to make information that is already accessible, though manually, more useable. This usability depends on its graph-database implementation.  To measure the value of this implementation through comparison, rather than perform human experiments, we reference an implementation where all information extracted from \bron's sources is stored in files using JSON formatting. These files are then read into memory and queried.  This is faster than manual querying.  We name \bron's implementation(which uses ArangoDB~\footnote{\url{http://arangodb.com}})  \brongdb , and the JSON-based implementation, \bronjson.
In this section we present the time and memory costs for \brongdb and draw comparisons between it and \bronjson for a set of small, medium and large queries.

\paragraph{Initialization and Coding}

%

The data within \brongdb is populated by a depth-first search of
unique nodes and edges. The time complexity is $O(|N| + |E|)$ where
$N$ denotes nodes, and $E$ denotes edges. Nodes, represent \TACTICS,
\TECHNIQS, \ATTACKPATTERNS, \WEAKNESS{es} and \VULNERABILITIES. Edges
link node entries linked in the sources.

\paragraph{Comparison Setup}

We used an Ubuntu 20.04 LTS instance on a private OpenStack cloud with
4 Intel Xeon CPU E5-2630 v4 2.20 GHz and 16 GB RAM. \bronjson uses 294 MB
of disk space, and \brongdb uses 351 MB of disk space.

Per Table~\ref{tab:query_results}, \textit{Query Time} is the time
taken between running the query and returning the query results using
the Python time module.  \textit{Query Memory} is the memory usage of
running the query using the Python memory profiler. We recorded the
execution statistics of three queries, each repeated 30 times. To
compare the queries, we display how many nodes in \bron accessed to
obtain the output of the query. In this count, nodes are not unique.

There are 3 queries. Two require searching graph paths. For them, the input is a CSV file of node IDs and the output is a CSV file with the IDs of nodes connected
to each of the input nodes along an edge in \bron.  For the third query, about riskiest software, the output is the \CONFIG with the highest sum of CVSS scores for \VULNERABILITIES connected to the \CONFIG.

\paragraph{Query Performance Results}

The results for the queries are summarized in Table~\ref{tab:query_results}.

\textbf{Threats connected to top 10 CVEs}. This query set is examined in
Section~\ref{sec:top10CVEs}.
\textit{Query Time}: The mean query time for \brongdb is nearly 4 times
faster than when querying \bronjson.

\textbf{Threats and vulnerabilities connected to top 25 CWEs}. This query set is examined in Section~\ref{sec:cwe-top-25}.
\textit{Query Time}: The mean query time for \brongdb is at least twice
as fast as that for \bronjson.

\textbf{Riskiest software}.
\textit{Query Time}: The mean query time for riskiest software using
\brongdb is nearly 10 seconds slower than when using \bronjson.

\begin{table}[tb]
  \small
  \centering
  \caption{Query performance results from 30 repeats}
  \label{tab:query_results}
  \begin{tabular}{lllll}
    \textbf{Query Measurement} & \textbf{Min} & \textbf{Max} & \textbf{Mean} & \textbf{SD}\\
    \hline \hline
\multicolumn{5}{c}{\textbf{Threats connected to top 10 CVEs (Small: 390 nodes)}} \\
\hline
{\bronjson  Time (sec)} & {12.29} & {18.46} & {15.62} & {1.35}\\
{\brongdb  Time (sec)} & {3.58} & {4.85} & {4.00} & {0.35}\\  \hline
{\bronjson  Memory (MB)} & {2.5e3} & {3.6e3} & {3.1e3} & {0.22}\\
{\brongdb  Memory (MB)} & {62} & {62} & {62} & {0.01}\\
\hline \hline
\multicolumn{5}{c} {\textbf{Threats and vulnerabilities for top 25 CWEs (Medium: 322K nodes)}} \\
\hline
{\bronjson  Time (sec)} & {64.71} & {72.54} & {67.69} & {2.15}\\
{\brongdb  Time (sec)} & {26.54} & {37.85} & {29.03} & {2.61}\\  \hline
{\bronjson  Memory (MB)} & {2.59e3} & {3.35e3} & {3.04e3} & {0.18e3}\\
{\brongdb  Memory (MB)} & {65.77} & {105.82} & {76.61} & {9.61}\\
\hline \hline
\multicolumn{5}{c} {\textbf{Riskiest software (Large 2,453K nodes)}} \\
\hline
{\bronjson  Time (sec)} & {14.25} & {24.21} & {19.45} & {1.70}\\
{\brongdb  Time (sec)} & {24.70} & {37.36} & {28.68} & {2.94}\\  \hline
{\bronjson  Memory (MB)} & {2.88e3} & {3.50e3} & {3.25e3} & {0.17e3}\\
{\brongdb  Memory (MB)} & {216.21} & {222.79} & {217.76} & {1.86}\\
\end{tabular}
\end{table}

\paragraph{Availability}

The code is available at
\url{https://github.com/ALFA-group/BRON}. Code and data for the
analysis are available in \texttt{Python} with tutorials in
\texttt{Jupyter} notebooks. There is also a public instance of \Bron
hosted in an ArangoDB that is updated daily at
\url{http://bron.alfa.csail.mit.edu:8529}.

\section{Related Work}
\label{sec:related-work}

The sources of \bron are variously described as  \textit{industry standard of common names} (CVE), a \textit{list} (CVE and CWE), an \textit{encyclopedia} (CWE), \textit{comprehensive dictionary and classification taxonomy} (CAPEC) and \textit{knowledge base} (ATT\&CK).  They have multiple counterparts, also with a variety of descriptors.  Among them, but not exclusively are MISP \cite{wagner_misp_2016}, CTI \cite{luatix}, OVM \cite{ovm}, FireEye OpenIOC \cite{ioc}, STIX \cite{stix}, IDS rules \cite{rouached_efficient_2012}, OpenC2 \cite{muoio}, UCO \cite{uco}, VERIS \cite{veris}, and IODEF \cite{rfc}.  We chose \bron's sources for their composite information value and their popularity. Public data sources have been analyzed previously. For example,
\cite{ozment2007vulnerability} studied and analyzed NVD, and pointed
out several limitations. \cite{bullough2017predicting} describe some
statistics of \VULNERABILITIES and \WEAKNESS. The \VULNERABILITIES and
CVSS was analyzed by~\cite{feutrill2018effect}. \bron contributes an
analysis over more public threat data sources.

Similar threat and vulnerability data within knowledge graphs and ontologies exist. For example, the STUCCO ontology and knowledge graph incorporates information from 13 structured data sources and provides relationships among 15 entity types including software, vulnerabilities, and attacks \cite{iannacone2015developing}. The SEPSES knowledge graph links information from standard data sources including CVEs, CAPECs, CWEs, CPEs, and CVSS for vulnerability scoring \cite{kiesling2019sepses}. SEPSES provides use cases for vulnerability assessment and intrusion detection with sample queries. SEPSES uses some of \bron's sources but \bron support a broader set of analyses and inform users of the extent of the data. This same claim holds for \url{www.cvedetails.com} which provides an easy to use web interface to CVE vulnerability data only. 

Threat modeling models the software system, potential attacker goals and techniques, and potential threats to the system so that cyber defenders can identify and mitigate vulnerable assets. There are several threat modeling methods that vary in comprehensiveness, abstraction, and focus. These methods include but are not limited to STRIDE, PASTA, LINDDUN, and Attack Trees~\cite{shevchenko2018threat}. \bron uses a single graph to relationally connect entries from sources ranging from tactics to vulnerable software. 


A community-based approach like \textit{The Githubification of InfoSec} can speed learning for defenders~\cite{githubification2019}. It has three components: Insight, Analytics, and Analysis. For Insight the MITRE ATT\&CK framework is suggested.  For analytics and analyses sharing analytic security data mining via \texttt{Jupyter} notebooks is proposed. On example is CALDERA Pathfinder~\cite{calderaPathfinder2020} that show what a vulnerability exposes to an adversary. It is arguable that \bron could enhance these approaches. In terms of insights, it includes more than ATT\&CK and unifies inter-source linking.  The \texttt{Python} scripts in \bron are lower level stack components.  \bron offers abstract knowledge linked to concrete knowledge.


\section{Discussion}
\label{sec:discussion}

\bron offers a more coherent access point to new data consumers of these sources. It lowers the entry bar for performing cross-referenced analyses and enables analysts to think more contextually about attacks and vulnerabilities. It enables both cyber-hunting and reactive, preventative and forensic security analysis on strategic, operational and tactical levels. However, there are several limitations to our data-driven approach and its support by \bron's graph.

First, \bron relies only publicly reported data from NVD, and numerous
vulnerabilities exist that do not have CVE IDs. Additionally the data has biases due to reporting by diverse sources with different interests and product offerings. All the data biases
of each individual source, exist in \bron. See~\cite{topcwe} for examples of data and metric bias.

Second, given \bron is reliant on the quality of the available public
data, a major threat to this work's validity is the integrity of the
same data. There could be curation errors such as inconsistent severity scores and we have uncovered gaps.  No public resource in this context will ever be complete. The sources are also sensitive to data aging and heterogeneity: not all products use the same versioning standard, and older products can accumulate newly disclosed vulnerabilities.  We tried to address these sensitivities in one instance when we considered only the most recently released version of a \config.
gaps. The risks for the data breaching confidentiality is low since the data is public. The threat to availability of data depends on the availability of the data sources we amalgamated.
Directly ascribable to \bron is  a risk that \bron's snapshot of the sources is out of date and misses new updates to them. This is mitigated by the simple creation script we have developed.

In terms of the findings of our inquiries, we found both ``local'' uneven availability in the data sources (when we looked at Top 10 and Top 25 Alerts) and aggregate uneven availability.  This places a caveat on the particular query responses we found.   The same public data is also used to build predictive models, so the caveat carries over to this domain. Examples of modeling include addressing situational awareness ~\cite{liu2020rule}, predicting missing edges between CVE, CWE and CAPEC ~\cite{xiao2019}, and investigating data breaches with semantic analysis of ATT\&CK ~\cite{noor2019machine}. \bron's  analysis of the comprehensiveness of the public sources could potentially be extended to inform the bias and performance validity of
predictive models.

\section{Future Work}\label{sec:future-work}

In future work, we plan analyses that take additional care to align comparisons along a date or age of product. CAPEC and CWE has information regarding similar entries that can be utilized as well. We have not studied data source entity similarity (connections), only similarity between data sources. In the future the CAPEC and CWE sources offer information regarding similar entries that can be leveraged. In addition, we can try to use \bron to predict missing connections.

\Bron can also potentially support threat modeling and network vulnerability analysis of existing networks. There is also scope for performance enhancements by storing \bron in a dedicated graph database format. Finally, the utility of \bron for practitioners needs to be further evaluated.

\section*{Acknowledgments}

This material is based upon work supported by the DARPA Advanced Research Project Agency (DARPA) and Space and Naval Warfare Systems Center, Pacific (SSC Pacific) under Contract No. N66001-18-C-4036

\bibliographystyle{ACM-Reference-Format}
\bibliography{bron_bridging_public_threat_data_for_cyber_hunting}


\end{document}